\documentclass[prd,preprintnumbers,twocolumn,eqsecnum,floatfix,a4paper,nofootinbib,superscriptaddress]{revtex4}
\usepackage{color}
\usepackage{calc}
\usepackage{amsmath,amssymb,graphicx}
\usepackage{amssymb,amsmath}
\usepackage{tensor}
\usepackage{bm}
\usepackage{microtype}
\usepackage{booktabs}
\usepackage{times}
\usepackage[varg]{txfonts}
\usepackage[colorlinks, pdfborder={0 0 0}]{hyperref}
\usepackage{subfigure}
\definecolor{LinkColor}{rgb}{0.75, 0, 0}
\definecolor{CiteColor}{rgb}{0, 0.5, 0.5}
\definecolor{UrlColor}{rgb}{0, 0, 0.75}
\hypersetup{linkcolor=LinkColor}
\hypersetup{citecolor=CiteColor}
\hypersetup{urlcolor=UrlColor}
\allowdisplaybreaks
\usepackage[utf8]{inputenc}
\usepackage{ulem}
\DeclareFontFamily{OT1}{pzc}{}
\DeclareFontShape{OT1}{pzc}{m}{it}{<-> s * [1.10] pzcmi7t}{}
\DeclareMathAlphabet{\mathpzc}{OT1}{pzc}{m}{it}

\newcommand{\sk}[1]{}

\newcommand{\Rmnum}[1]{\expandafter\@slowromancap\romannumeral #1@}
\makeatother

\newcommand{\bt}[1]{\mathbf{#1}}

\def\mc{\mathcal}

\def\be{\begin{equation}}
\def\ee{\end{equation}}
\def\bea{\begin{eqnarray}}
\def\eea{\end{eqnarray}}
\def\bn{\begin{enumerate}}
\def\en{\end{enumerate}}
\def\bsube{\begin{subequations}}
\def\esube{\end{subequations}}

\def\Pr{{P}}
\def\vtheta{{\bm \theta}}
\def\H{{\mc{H}}}
\def\Hl{{\mc{H}_\textsc{l}}}
\def\Hu{{\mc{H}_\textsc{u}}}
\def\Zl{{\mc{Z}_\textsc{l}}}
\def\Zu{{\mc{Z}_\textsc{u}}}
\def\Olu{{\mc{O}_\textsc{u}^\textsc{l}}}
\def\Blu{{\mc{B}_\textsc{u}^\textsc{l}}}
\def\Rlu{{\mc{R}_\textsc{u}^\textsc{l}}}
\def\plu{{\mc{P}_\textsc{u}^\textsc{l}}}
\def\mz{{m^z}}
\def\Dls{{D_\textsc{ls}}}
\def\Dl{{D_\textsc{l}}}
\def\Ds{{D_\textsc{s}}}
\def\zl{{z_\textsc{l}}}
\def\zs{{z_\textsc{s}}}

\def\Dlc{{D_\textsc{l}^\textsc{c}}}
\def\Dsc{{D_\textsc{s}^\textsc{c}}}
\def\Dc{{D^\textsc{c}}}

\begin{document}
	
\preprint{LIGO-P1800155-v2}
	
\title{Identifying strongly lensed gravitational wave signals from binary black hole mergers} 

\author{K. Haris}
\author{Ajit Kumar Mehta}
\author{Sumit Kumar}
\affiliation{International Centre for Theoretical Sciences, Tata Institute of Fundamental Research, Bangalore 560089, India}
\author{Tejaswi~Venumadhav}
\affiliation{School of Natural Sciences, Institute for Advanced Study, Princeton, New Jersey 08540, USA}
\author{Parameswaran~Ajith}
\affiliation{International Centre for Theoretical Sciences, Tata Institute of Fundamental Research, Bangalore 560089, India}
\affiliation{Canadian Institute for Advanced Research, CIFAR Azrieli Global Scholar, MaRS Centre, West Tower, 661 University Ave., Suite 505, Toronto, ON M5G 1M1, Canada}

\date{\today}
\begin{abstract}
Based on the rate of gravitational-wave (GW) detections by Advanced LIGO and Virgo, we expect these detectors to observe hundreds of binary black hole mergers as they achieve their design sensitivities (within a few years). A small fraction of them can undergo strong gravitational lensing by intervening galaxies, resulting in multiple images of the same signal. To a very good approximation, the lensing magnifies/de-magnifies these GW signals without affecting their frequency profiles. We develop a Bayesian inference technique to identify pairs of strongly lensed images among hundreds of binary black hole events, and demonstrate its performance using simulated GW observations. 
	\end{abstract}
	
	
	\maketitle
	
\section{Introduction}
Arthur Eddington’s 1919 observation of the gravitational bending of light was the first observational test that heralded the remarkable success of general relativity (GR)~\cite{1920RSPTA.220..291D}. Recent observations of gravitational waves (GWs) by LIGO~\cite{TheLIGOScientific:2014jea} and Virgo~\cite{TheVirgo:2014hva} have vindicated one of the most famous astrophysical predictions of GR~\cite{gw150914, gw151226, gw170104, gw170608, gw170814, gw170817}. While gravitational lensing (of electromagnetic waves) has been well established as a powerful astronomical tool (see, e.g.,~\cite{Bartelmann:2010fz} for a review), GW observations are opening up an emerging branch of observational astronomy (see, e.g.,~\cite{Sathyaprakash:2009xs} for a review). 

GWs are gravitationally lensed by intervening mass concentrations along the line of sight from the source to the observer, in a manner similar to electromagnetic waves. Several previous papers in the literature have considered the resulting phenomenology for GWs from a variety of compact object mergers~\cite{1996PhRvL..77.2875W,2003ApJ...595.1039T,2004A&A...423..787T,2004PhRvD..69b2002S,2010PhRvL.105y1101S,2011MNRAS.415.2773S,2013JCAP...10..022P,2014JCAP...10..080B,2017arXiv170204724D,2015JCAP...12..006D}. Recent estimates of the lensing rates have shown that at upgraded sensitivities of Advanced LIGO, a small fraction ($< 1\%$) of the detected GW signals from stellar--mass binary black hole mergers  can be strongly lensed by intervening galaxies and clusters~(see, e.g., \cite{Ng:2017yiu}). These mergers would produce multiple ``images'' at different times, with significantly different intrinsic masses and redshifts~\cite{2017PhRvD..95d4011D,Ng:2017yiu,2018MNRAS.475.3823S,2018MNRAS.476.2220L}. It has even been suggested that a significant fraction of the detected merger population was strongly lensed~\cite{2018arXiv180205273B}, which would require a strong redshift evolution of the intrinsic merger rate. In the standard case, the lensed fraction is expected to be small, but LIGO and Virgo are expected to detect hundreds of binary black hole mergers over the next few years~\cite{Abbott:2016nhf}; thus it is quite likely that some of the detected signals will be strongly lensed. Identification of strongly lensed GW signals would be rewarding. On the one hand, we will be verifying a fundamental prediction of GR using a messenger entirely different from electromagnetic radiation \cite{Liao:2017ioi,Fan:2016swi}. In addition, such a detection can potentially enable astrophysical studies of the lens galaxy and the host galaxy \cite{Smith:2018kbc}.

In this paper, we consider the problem of observationally identifying a pair of lensed signals coming from a single merger among hundreds of unrelated merger signals. From the perspective of the observer, these lensed images would appear as different GW signals that are separated by time delays of minutes to weeks. The observed gravitational waveform depends on the zenith angle and the azimuth of the merger relative to the detectors, which will be different for each image. Moreover, each image will be observed against a different realization of the detector noise. This makes it difficult to compare multiple images at the waveform level, and necessitates a comparison in the space of the estimated intrinsic parameters.  

We work in the geometric optics limit, which applies when the wavelength of the GW signal is small compared to the Schwarzschild radius of the lens mass ($\lambda_\mathrm{GW} \ll  2 G M_\mathrm{lens}/c^2$). This approximation can fail to model the lensing of GW signals from supermassive black holes lensed by intervening supermassive black holes or dark matter halos with masses $\sim 10^8 M_\odot$ (which leads to interesting wave effects that could be observed by LISA~\cite{2003ApJ...595.1039T,2004A&A...423..787T}), or of GW signals from stellar mass black holes lensed by intermediate mass black holes or compact halo objects with masses $\sim 10^3 M_\odot$ (which can lead to interesting wave effects observable by LIGO~\cite{Lai:2018rto,Jung:2017flg}). However, the geometric optics approximation is adequate to model the GW signals from stellar--mass black holes observed in LIGO/Virgo that are lensed by galaxies. In this regime, lensing will magnify/de-magnify the GW signal without affecting its shape. Since the parameters of the merging binary are estimated  by comparing the data with theoretical templates of the expected signals~(see, e.g,~\cite{TheLIGOScientific:2016wfe}), the estimated parameters (barring the estimated luminosity distance, which is degenerate with the magnification and hence will be biased) of these different signals will be mutually consistent. 

We develop a Bayesian formalism for identifying strongly lensed and multiply imaged GW signals from binary black hole merger events among hundreds of unrelated merger signals. From each pair of GW signals, we compute the Bayesian odds ratio between two hypotheses: 1) that they are the lensed images of the same merger event, 2) that they are two unrelated events. Using simulated GW events (lensed as well as unlensed), we show that this odds ratio is a powerful discriminator that will allow us to identify strongly lensed signals. Our method can be easily integrated with the standard Bayesian parameter estimation pipelines that are used to analyze LIGO and Virgo data~\cite{Veitch:2014wba}.

The paper is organized as follows: Section~\ref{sec:lensing} is a brief primer on gravitational lensing. In Sec.~\ref{bayes_facotr_formulation}, we develop a Bayesian odds ratio between the two hypotheses (lensing and null). Using simulated GW observations, we test the efficacy of this odds ratio in distinguishing pairs of lensed GW signals from pairs of unlensed signals in Sec.~\ref{sec:astro_sim}. Finally, we present some conclusions and comment on future directions in Sec.~\ref{sec:conclusions}. A detailed description of our astrophysical simulation of lensed GW merger events is presented in Appendix~\ref{lens_simulation}. 

\section{A gravitational lensing primer}
\label{sec:lensing}

Gravitational lensing describes the effect of mass inhomogeneities along the line of sight on the propagation of radiation between a source and an observer~\cite{1967ApJ...150..737G}. The terminology of \emph{strong} gravitational lensing is used when the dominant effect is due to only a few discrete mass aggregations along the line of sight. For sources at moderate redshifts in typical cosmologies, the strong lensing probability (the so-called \emph{optical depth} $\tau_{\rm S}$) is small~\cite{2007MNRAS.382..121H,2011ApJ...742...15T}, and hence the most frequently studied case involves a single mass concentration (the ``single-lens-plane'' case~\cite{1992grle.book.....S}). 

In the single-lens-plane case, the radiation propagates on geodesics of the background spacetime between the source- and the lens planes, and the lens-plane and the observer. The effect of the lens is described by the dimensionless Fermat potential $\phi(\bt x, \bt y)$, where $\bt x$ and $\bt y$ are angular coordinates on the lens- and source planes, respectively. The potential $\phi(\bt x, \bt y)$ is the scaled time-delay due to the geometrical path length, and the gravitational potential of the deflecting mass. 

Let us consider a lens with a surface mass density profile $\Sigma(\bt x)$. For a source at an angular location $\bt y$ on the source plane, the Fermat potential takes the form
\begin{equation}
\phi(\bt x, \bt y)  = \frac12 (\bt x - \bt y)^2 - \psi(\bt x),  \label{eq:fermatphi} 
\end{equation}
where 
\begin{equation}
\psi(\bt x)  = \frac1\pi \int d^2 \bt x^\prime \, \kappa(\bt x^\prime) \ln{\vert \bt x - \bt x^\prime \vert}, \label{eq:deflectionpotential} \\
\end{equation}
with $\kappa(\bt x)  = {\Sigma(\bt x)}/{\Sigma_{\rm cr}}$, where the critical density is given by 
\begin{equation}
{\Sigma_{\rm cr}}  = \frac{c^2 \Ds}{4\pi G \Dl \Dls}. 
\end{equation}
Above,  $\Ds, \Dl,$ and $\Dls$ are the angular diameter distances between the observer and the source, the observer and the lens, and the lens and the source, respectively.

Under the geometrical optics (i.e., the short wavelength) approximation, a source at a location $\bt y$ has discrete images at extrema of the Fermat potential $\phi(\bt x, \bt y)$ on the lens- (or the image-) plane. From Eq.~\eqref{eq:fermatphi}, the image-locations $\bt x$ satisfy the lens equation
\begin{equation}
\bt y  = \bt x - \bm{\alpha}(\bt x),\label{eq:lenseq} 
\end{equation}
where 
\begin{equation}
\bm{\alpha}(\bt x)  = \bm\nabla \psi(\bt x). \label{eq:deflection}
\end{equation}
In practice, Eq.~\eqref{eq:lenseq} is an implicit equation that must be inverted to obtain the image-locations. Note that given the surface-mass density profile $\kappa(\bt x)$, the deflection angle $\bm{\alpha}(\bt x)$ is completely specified using Eqs.~\eqref{eq:deflection} and \eqref{eq:deflectionpotential}. The geometrical magnification factor $\mu(\bt x_i)$ of each image $i$ is given by the inverse of the determinant of the lensing Jacobian matrix $d{\bt y}/d{\bt x}$ evaluated at $\bt x_i$. We compute the (proper) mutual time delay between two images at $\bt x_1$ and $\bt x_2$ as seen by an observer using the Fermat potential as follows:
\begin{align}
c \Delta t_{12} (\bt y) = (1+\zl) \frac{\Ds \Dl}{\Dls} \left[ \phi(\bt x_1, \bt y) - \phi(\bt x_2, \bt y) \right], \label{eq:timedelay}
\end{align}
where $\zl$ is the redshift of the lens. We use the magnification $\mu_i$ of the images and time delay $\Delta t_{12}$ between them to modify the GW signal, as described in detail in Appendix~\ref{lens_simulation}. In our application, we assume that the surface-mass profiles of the lenses have the simple singular isothermal ellipsoid (SIE) form, for which we can analytically calculate the deflection angle, magnification, and Fermat potential at any given image-plane location~\cite{1994A&A...284..285K}. More information is provided in Appendix~\ref{lens_simulation}.
\section{Bayesian model selection of strongly lensed GW signals from binary black hole mergers} 
\label{bayes_facotr_formulation}
\begin{figure*}[tbh] \begin{center}
		\includegraphics[width=3.4in]{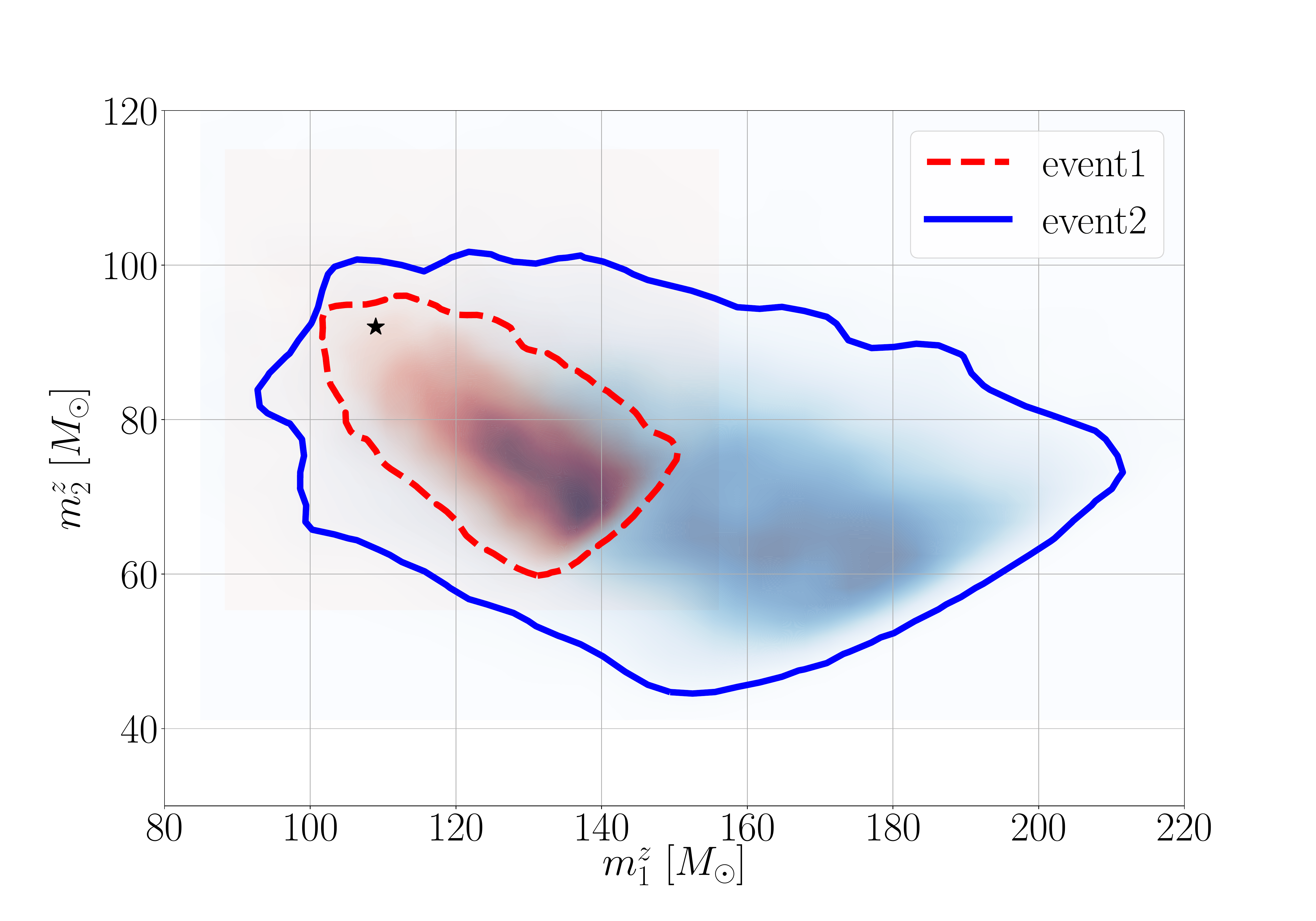}
		\includegraphics[width=3.4in]{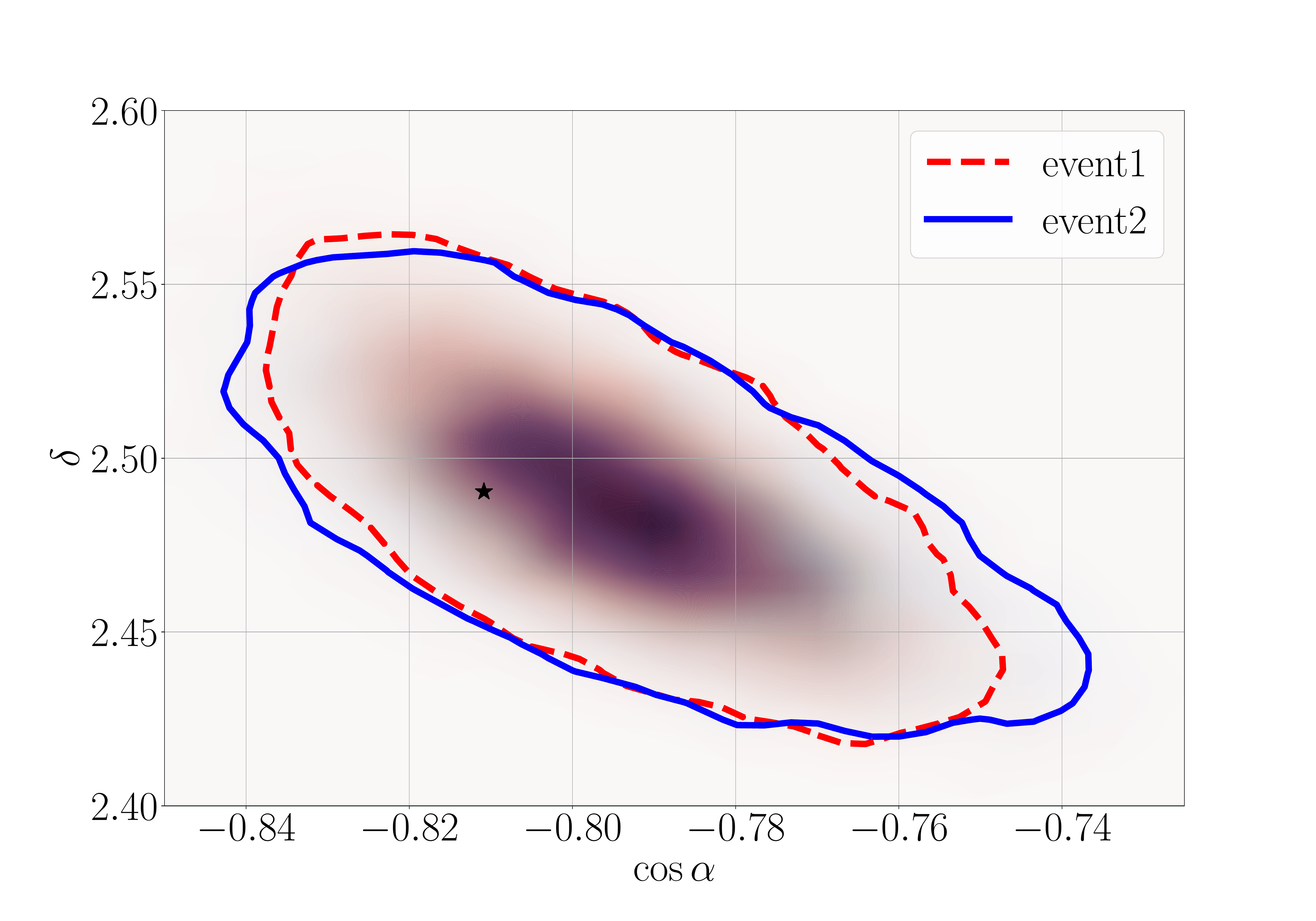}
		\caption{95\% credible regions of the marginalized posteriors of the redshifted masses  $m_1^z, m_2^z$ (left) and sky location $\cos \alpha, \delta$ (right) of lensed images of a sample binary black hole merger event. Black stars show the actual injected parameters.}
		\label{fig:Posterior_sample_plots}
\end{center} \end{figure*}

%
Consider a data stream $d(t)$ of a GW detector containing a signal $h(t, \vtheta)$ described by a set of parameters $\vtheta$ and some stochastic noise $n(t)$: 
\begin{equation}
d(t) = n(t) + h(t, \vtheta). 
\end{equation}
For binary black holes in quasi-circular orbits, the GW signals $h(t, \vtheta)$ are described by a set of parameters $\vtheta$ that consists of the redshifted masses $(\mz_1, \mz_2)$, the dimensionless spin vectors $(\bm \chi_1, \bm \chi_2)$, the time of coalescence $t_0$ and the phase at coalescence $\varphi_0$, sky location $(\alpha, \delta)$, the inclination $\iota$ of the binary, the polarization angle $\psi$ and the luminosity distance $d_L$ to the source. The posterior distribution of the set of parameters $\vtheta$ can be computed from the data using the Bayes theorem as follows:
\begin{equation}
\Pr(\vtheta | d) = \frac{ \Pr(\vtheta ) \, \Pr(d|\vtheta)}{\Pr(d)}  \, ,
\label{eq:bayes_theorem}
\end{equation}
where $\Pr(\vtheta)$ denotes the prior distribution of $\vtheta$, $\Pr(d|\vtheta)$ is the likelihood of the data $d$ assuming the signal $h(t,\vtheta)$ and 
\be
\Pr(d ) := \int d\vtheta \, \Pr(\vtheta) \, \Pr(d|\vtheta)
\label{eq:marg_likelihood_single_event}
\ee 
is called the marginalized likelihood. If  $n(t)$ can be well approximated by a stationary Gaussian process with mean zero and a one-sided power spectral density $S_n(f)$, then the likelihood is given by 
\begin{equation}
\Pr(d|\vtheta) = \mathcal{N} ~ \exp \left\{ - \frac{1}{2}  \langle d - h ~|~ d - h \rangle \right\},
\end{equation}
where $\mathcal{N}$ is a normalization constant and $\langle . | . \rangle$ denotes the following noise-weighted inner product:
\begin{equation}
\langle a | b \rangle := 2 \int_{f_\mathrm{low}} ^{f_\mathrm{upp}} df \, \frac{\tilde{a}^*(f)\tilde{b}(f) + \tilde{a}(f)\tilde{b}^*(f)}{S_n(f)}.
\label{eq:nwip}
\end{equation}
Above, $f_\mathrm{low}$ and ${f_\mathrm{upp}}$ denote the lower and upper cutoff frequencies of the detector's bandwidth, $\tilde{a}(f)$ denotes the Fourier transform of $a(t)$ and a $^*$ denotes complex conjugation.

If we have two data streams $d_1$ and $d_2$ containing GW signals from binary black holes, there is a small probability that these signals are lensed versions of a single merger event. In the geometric optics approximation, lensing does not affect the frequency profile of the signal. As a result, the lensed signals would correspond to the same set of parameters $\vtheta$ (except the estimated luminosity distance, which will be biased due to the unknown magnification). In order to determine whether $d_1$ and $d_2$ contain lensed signals from the same binary black hole merger, we compute the \emph{odds} ratio between two hypotheses: 
\begin{itemize}
\item $\Hl$: The data set $\{d_1, d_2\}$ contain lensed signals from a single binary black hole merger event with parameters $\vtheta_1 = \vtheta_2 = \vtheta$.
\item $\Hu$: The data set $\{d_1, d_2\}$ contain signals from two independent binary black hole merger events with parameters $\vtheta_1$ and $\vtheta_2$.
\end{itemize}
The odds ratio between $\Hl$ and $\Hu$ is the ratio of the posterior probabilities of the two hypotheses. That is, 
\be
\Olu = \frac{\Pr(\Hl|\{d_1,d_2\})}{\Pr(\Hu|\{d_1,d_2\})}~,
\ee
Using Bayes theorem we can rewrite the odds ratio as
\be
\Olu =   \frac{\Pr(\Hl)}{\Pr(\Hu)} ~ \frac{\Pr(\{d_1,d_2\}|\Hl)}{\Pr(\{d_1,d_2\}|\Hu)} = \plu~ \Blu 
\label{OddsRatio_1}
\ee
Here $\plu := \frac{\Pr(\Hl)}{\Pr(\Hu)}$ is the ratio of prior odds of the two hypotheses while the Bayes factor $\Blu := \Zl/\Zu$ is the ratio of the marginalized likelihoods, where the marginal likelihood of the hypothesis $A$ is $\mathcal{Z}_A :=  \Pr(\{d_1,d_2\}|\mc{H}_A)$ with $A \in \{\textsc{l, u}\}$. Under the assumption of $d_1$ and $d_2$ being independent, the marginal likelihood of the ``null'' hypothesis equals the product of the marginal likelihoods from individual events, i.e.,  
\be 
\Zu = \Pr(d_1) \, \Pr(d_2), 
\label{eq:evidence_unlensed}
\ee 
where $\Pr(d_i)$ is the marginal likelihood from event $i$, defined in Eq.~(\ref{eq:marg_likelihood_single_event}). Now, we rewrite the marginal likelihood of the lensing hypothesis in terms of the likelihoods of $d_1$ and $d_2$ as
\bea
\Zl = \int d \vtheta ~\Pr(\vtheta)~\Pr(d_1|\vtheta)~ \Pr(d_2|\vtheta)~. \label{Evidence_1}
\eea
Using Eq.~(\ref{eq:bayes_theorem}), we can rewrite this as 
\be
\Zl = \Pr(d_1) \, \Pr(d_2) \, \int d\vtheta ~ \frac{\Pr(\vtheta | d_1) \, \Pr(\vtheta | d_2)}{\Pr(\vtheta)}
\label{eq:evidence_lensed}
\ee 
Combining Eqs.~(\ref{eq:evidence_unlensed}) and (\ref{eq:evidence_lensed}), we obtain the following expression for the Bayes factor:
\be
\Blu := \frac{\Zl}{\Zu} = \int d \vtheta~ \frac{\Pr(\vtheta|d_1)~\Pr(\vtheta|d_2)}{\Pr(\vtheta)}~.
\label{eq:lensing_bayes_factor}
\ee
Thus, the Bayes factor is the inner product of the two posteriors that is inversely weighted by the prior. This has an intuitive explanation: if $d_1$ and $d_2$ correspond to lensed signals from a single binary black hole merger, the estimated posteriors on $\vtheta$ would have a larger overlap, favoring the lensing hypothesis (see, e.g., Fig.~\ref{fig:Posterior_sample_plots}). The inverse weighting by the prior helps to down-weight the contribution to the inner product from regions in the parameter space that are strongly supported by the prior. The large overlap of the posteriors here is less likely to be due to the lensing but more likely due to the larger prior support to the individual posteriors. 

\begin{figure} [tb]\begin{center}
			\includegraphics[width=3.6in]{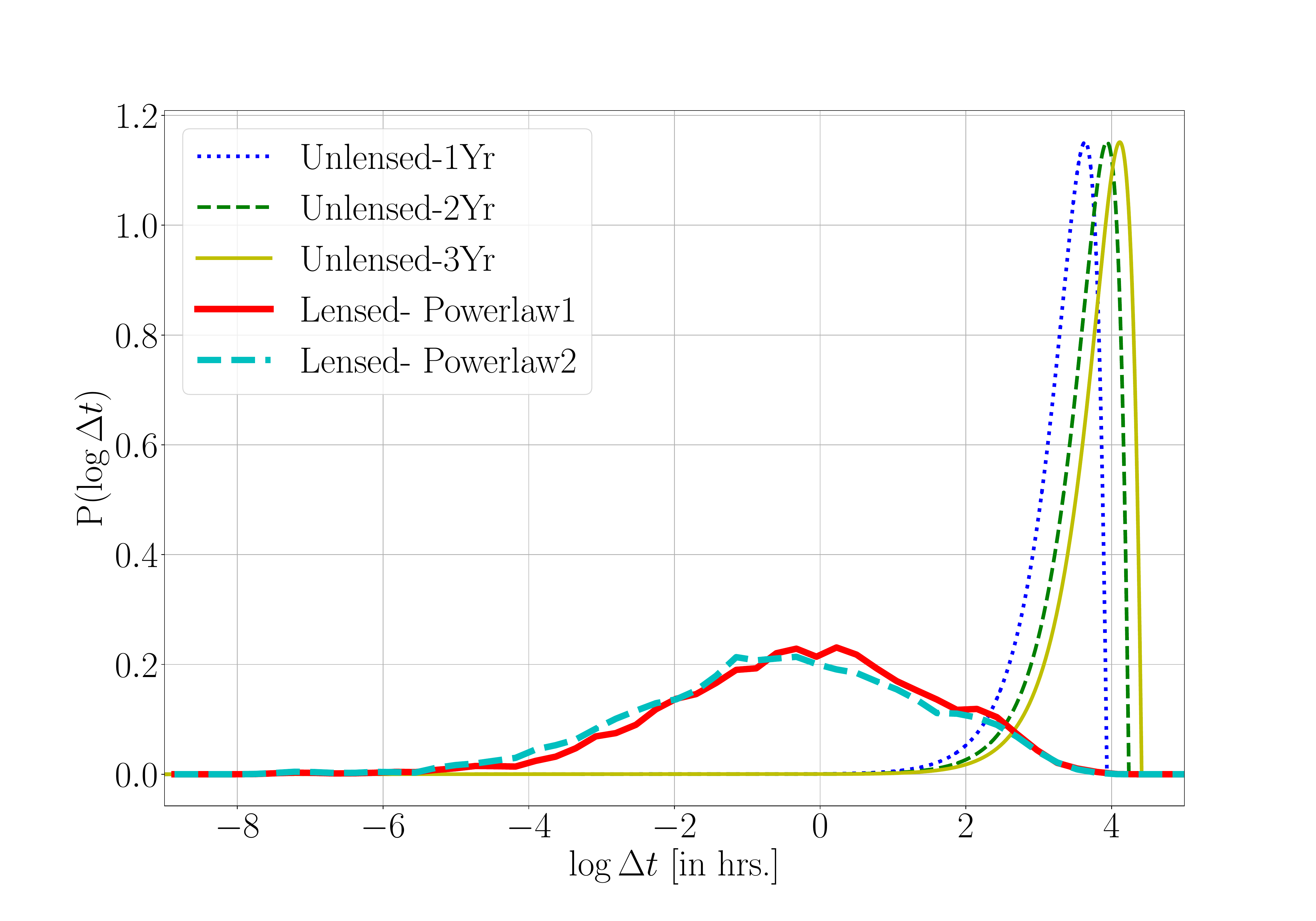}
			\caption{Distribution of the log of the time delay between lensed event pairs detected by the Advanced LIGO-Virgo network, along with the distribution from unlensed event pairs. The simulated binary black hole populations have their component masses (source-frame) distributed according to two power laws (see text); however, note that the time delays are practically insensitive to the specific form of the mass distribution. The  redshifts of the mergers are sampled with the distribution obtained in \cite{Dominik:2013tma}. We consider strong lensing produced by intervening galaxies. In order to compute the distribution of the time delay between \emph{unlensed} events, we assume that they follow a Poisson distribution with a rate of 10 mergers per month. The time delay distributions of unlensed event pairs get skewed towards larger values as we increase the observation time.} 
			\label{fig:Time_delay_hist}
		\end{center} \end{figure}

While the odds ratio developed above checks for the consistency between the estimated parameters of two GW signals, the time delay between them can also be used to develop a potential discriminator between lensed and unlensed events. This however, would require certain assumptions on the distribution of lenses (i.e., galaxies) and the rate of binary mergers. If we assume that binary merger events follow a Poisson process with a rate of $n$ events per month, one can compute the prior distribution $\Pr(\Delta t|\Hu)$ of time delay between pairs of unlensed events (see Fig.~\ref{fig:Time_delay_hist}). The prior distribution of the time delay between strongly lensed signals, $\Pr(\Delta t|\Hl)$,  would have a qualitatively different distribution, which can be computed using a reasonable distribution of the galaxies and a model of the compact binary mergers (see Sec.~\ref{sec:astro_sim} for details). Following Eq.\eqref{eq:marg_likelihood_single_event}, the marginal likelihood for the lensed/unlensed hypothesis can be computed from the time delay between two events $d_1$ and $d_2$ as 
\begin{equation}
\Pr_{\Delta t}(\{d_1, d_2\}|\H_A) = \int d\Delta t \, \Pr(\Delta t|\H_A) \, P(\{d_1, d_2\} | \Delta t, \H_A),
\end{equation}
where $A \in \{\textsc{l, u}\}$. Typical statistical errors in estimating the time of arrival of a GW signal at a detector are of the order of milliseconds --- much smaller than the typical time delay between any pair of events. Thus, the likelihood function $P_{\Delta t}(\{d_1, d_2\} | \Delta t, \H_A)$ of the time delay can be well approximated by a Dirac delta function at the true value $\Delta t_0$. Thus, the Bayes factor between the lensed and unlensed hypotheses can be written as 
\be
\Rlu = \frac{\Pr(\Delta t_0 |\Hl)}{\Pr(\Delta t_0|\Hu)}~,
\label{eq:bayesfactor_lensing_timedel}
\ee
where $\Pr(\Delta t_0 |\H_A)$ with $A \in \{\textsc{l, u}\}$ is the prior distribution of $\Delta t$ (under lensed or unlensed hypothesis) evaluated at $\Delta t = \Delta t_0$. The prior distributions are shown in Fig.~\ref{fig:Time_delay_hist}.

		\begin{figure}[tbh] \begin{center}
				\includegraphics[width=3.6in]{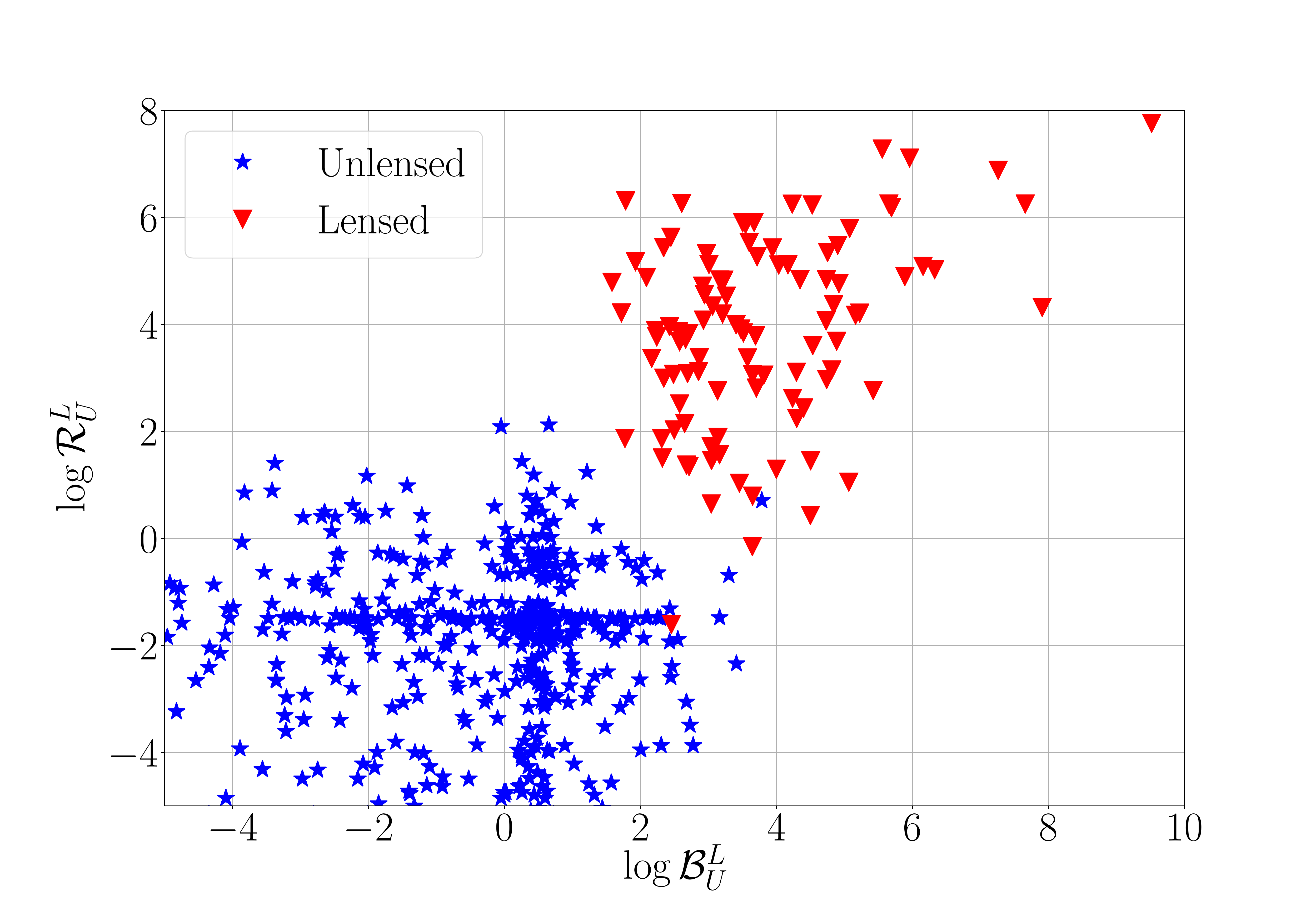}
				\caption{Scatter plot of the two Bayes factors $\Blu$ and $\Rlu$ computed from the unlensed (blue stars) and lensed (red triangles) event pairs. The Bayes factors computed from the posterior distribution of the binary's parameters ($\Blu$) and that computed from the time delay distribution ($\Rlu$) are in general correlated. However, they can be combined to improve our ability to distinguish lensed pairs from unlensed pairs. In this simulation, the component masses are distributed according to the second power law given in the text.}
				\label{fig:Bul_Bdeltat_scatter_dominik_p2_5yr}
			\end{center} \end{figure}

The Bayes factors $\Blu$ and $\Rlu$ could be combined to improve the discriminatory power between lensed and unlensed events. Figure \ref{fig:Bul_Bdeltat_scatter_dominik_p2_5yr} shows a scatter plot of $\Blu$ and $\Rlu$ computed from simulated pairs of lensed and unlensed events. As one can see, combining $\Blu$ and $\Rlu$ improves the discriminatory power. Note that, since the fraction of binary black hole mergers that are expected to produce strongly lensed signals is very small, the ratio of prior odds $\plu$ is a small number ($ < 1\%$). Hence, we need large values for the Bayes factors to confidently identify strongly lensed pairs of signals.   

\section{Testing the model selection}
\label{sec:astro_sim}

\begin{figure*}[tbh] \begin{center}
		\includegraphics[width=3.4in]{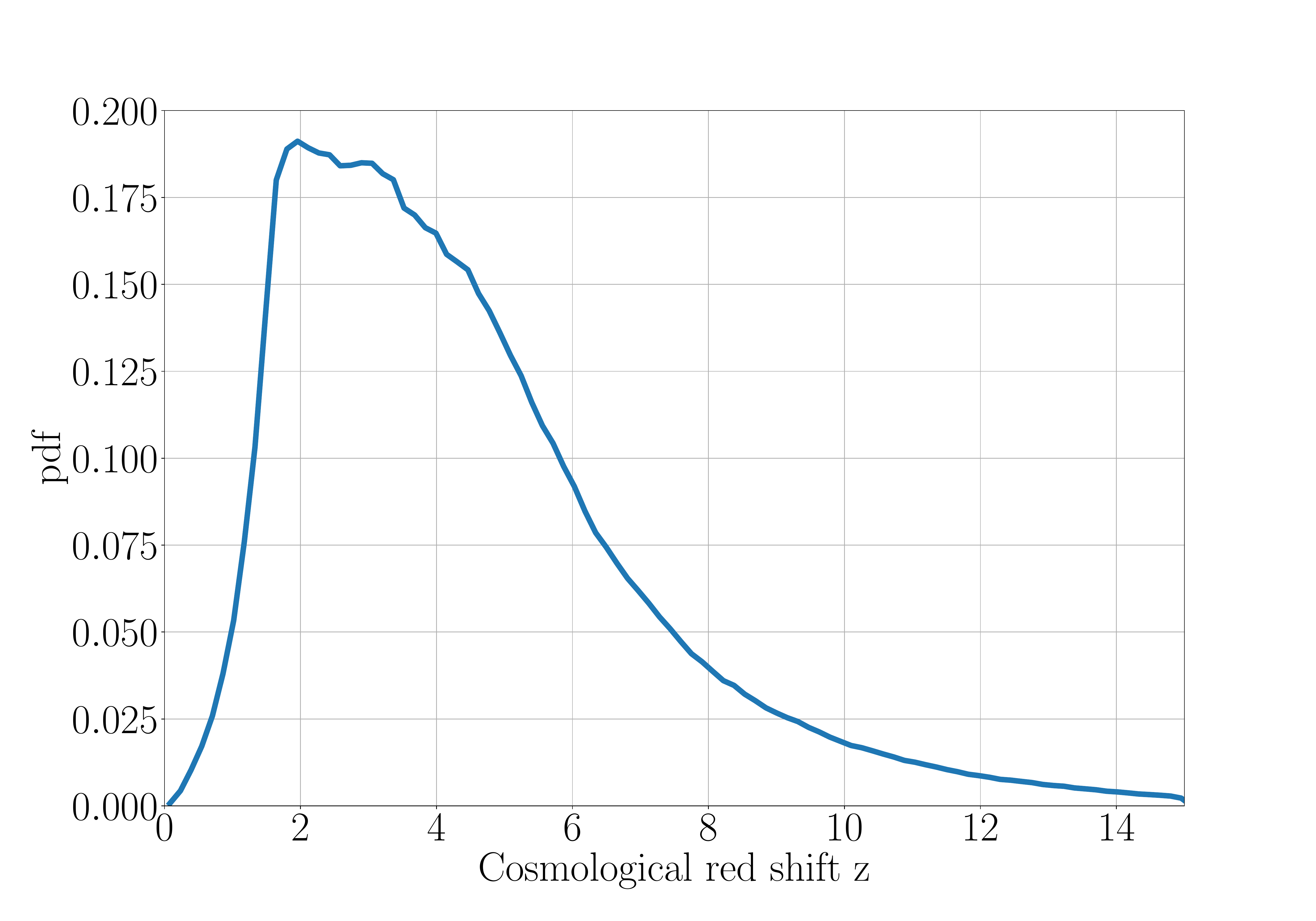}
		\includegraphics[width=3.4in]{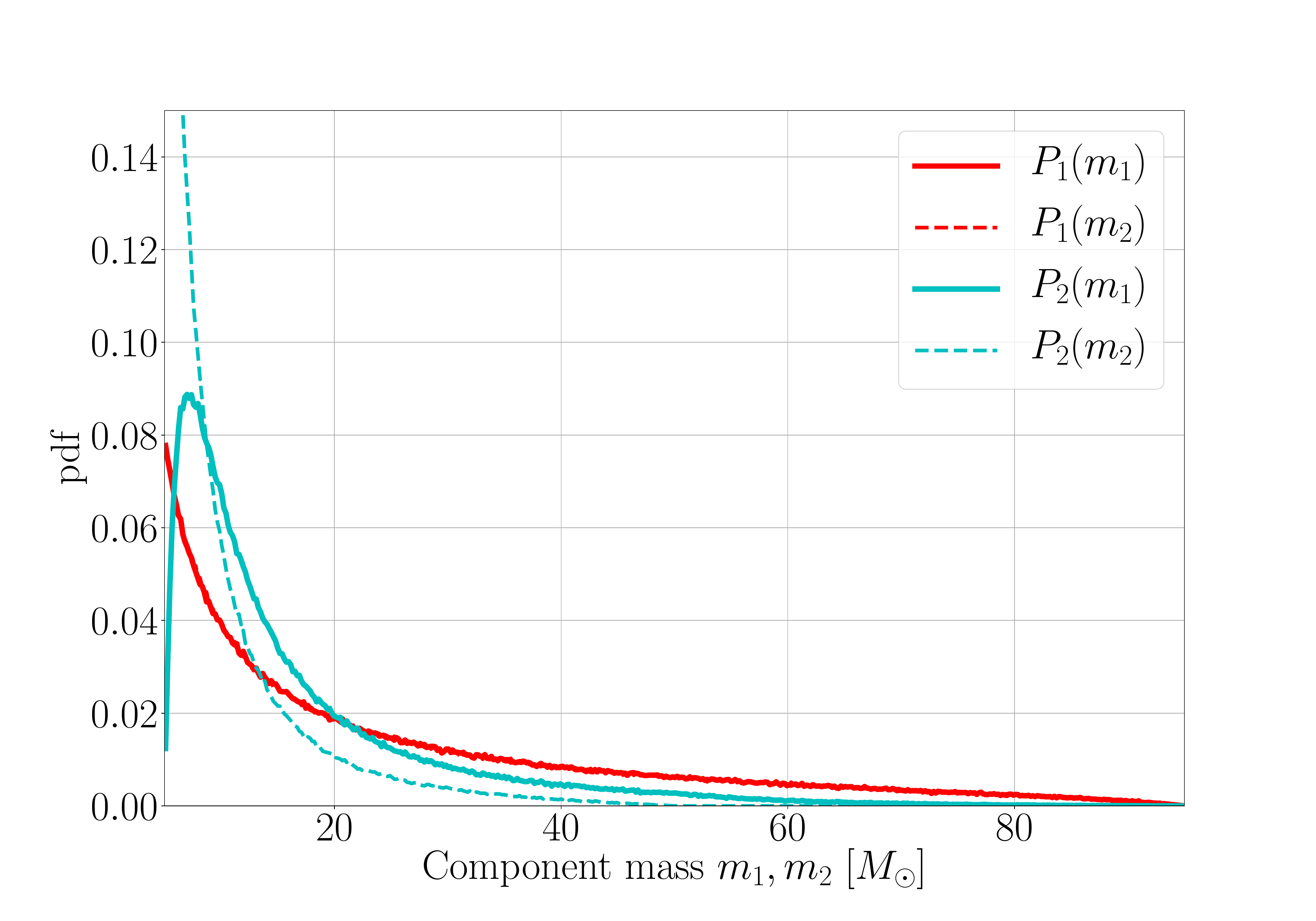}
		\caption{Probability distributions of the cosmological redshift (left) and component masses $m_1, m_2$ in the source frame (right) of the simulated binary black hole merger events.} 
		\label{fig:Dominik_z_powerlaw_m_hist}
\end{center} \end{figure*}

In this section we test the efficacy of our Bayesian model selection method to identify strongly lensed GW signals from binary black hole merger events. We simulate a population of coalescing binary black holes and compute the effect of strong lensing on the GW signals that they radiate. The binary black hole mergers are distributed according to the cosmological redshift distribution given in  \cite{Dominik:2013tma}. We use two different mass distributions proposed in \cite{Abbott:2016nhf} to sample component black hole masses $m_1$ and $m_2$:
\begin{enumerate}
\item Masses following a power-law $\Pr_1(m_1,m_2) \sim \frac{1}{m_1}\frac{1}{m_2}$ with $m_1, m_2 \geq 5 M_\odot$ and $m_1 + m_2  \leq 100 M_\odot$. 
\item Masses following a power-law $\Pr_2(m_1) = m_1^{-2.35}$ on the mass of the larger black hole, with the smaller mass distributed uniformly in mass ratio $m_1/m_2$ and with $5 M_\odot \leq m_1 + m_2  \leq 100 M_\odot$. 
\end{enumerate}
Figure \ref{fig:Dominik_z_powerlaw_m_hist} shows the redshift and mass distributions of the injections. The spin magnitudes $\chi_1 := ||\bm \chi_i||$ of component black holes are distributed uniformly between $0.$ and $0.99$, with random directions with respect to the orbital angular momentum.  The binaries are distributed uniformly in the sky (i.e., uniform in $\cos \alpha$ and $\delta$), and the inclination and polarization angles are sampled uniformly from polarization sphere (i.e., uniform in $\cos \iota$ and $\psi$). Note that the GW signals will be redshifted due to the cosmological redshift, and we infer the redshifted masses $m_{1,2}^z := m_{1,2} (1+z)$ through parameter estimation. 

Multiple images dominantly arise due to galaxy lenses~\cite{1991MNRAS.253...99F}. We assume that the galaxy lenses are well modeled by singular isothermal ellipses \cite{1991MNRAS.253...99F,1994A&A...284..285K}. The lens parameters, namely velocity dispersion $\sigma$ and axis-ratio $q$, are sampled from distributions modeled from the SDSS population of galaxies~\cite{0004-637X-811-1-20}. A detailed account on the lensing probability, sampling of lens galaxies and computation of the magnification factor and time delays is provided in Appendix~\ref{lens_simulation}. We simulate two populations of GW signals:
\begin{itemize}
\item \emph{Lensed:} Pairs of events with same parameters $\vtheta$, with parameter distributions as described above. We apply the lensing magnifications and time delays according to the prescription given in Appendix~\ref{lens_simulation}. 
\item \emph{Unlensed:} Pairs of events with random parameters $\vtheta_1$ and $\vtheta_2$, with parameter distributions as described above.
\end{itemize}
Figure~\ref{fig:Time_delay_hist} shows the distribution of time delays between pairs of lensed events as well as pairs of unlensed events from simulations assuming different distributions of source parameters. In the case of unlensed events, we compute the distribution of time delay assuming that the events follow a Poisson process with a rate of $n = 10$ events per month. Naturally the distribution of time delays between event pairs will depend only on the total observation time. The figure shows the time delay distributions from all pairs of events assuming observational runs of 1, 2 and 3 year duration.   

To simulate GW observation coming from each population, we inject simulated GW signals from binary black holes in colored Gaussian noise with the design power spectrum of the three-detector Advanced LIGO-Virgo network~\cite{aLIGOSensitivity,aLIGOSensitivity_reference, AvirgoBaseline}. The signals are modelled by the \textsc{IMRPhenomPv2} waveform family~\cite{Hannam:2013oca,Husa_2016IMRPhenomD,Khan_2016IMRPhenomD} which describes GW signals from the inspiral, merger and ringdown of binary black holes with precessing spins in quasi-circular orbits\footnote{Note that, in this waveform, the spin effects modeled in terms of two effective spin parameters~\cite{Hannam:2013oca,Ajith:2009bn}.}.

From simulated events that cross a network signal-to-noise ratio (SNR) threshold of 8, we estimate the posterior distributions of the parameters using the \textsc{LALInferenceNest} code~\cite{Veitch:2014wba}. This code provides an implementation of the Nested Sampling algorithm~\cite{skilling2006} in the \textsc{LALInference} software package of the  LIGO Algorithm Library \textsc{LALSuite}~\cite{Lalsuite}. From each population of injections (lensed and unlensed), we draw random pairs from the simulated events and compute the Bayes factor $\Blu$ defined Eq.~(\ref{eq:lensing_bayes_factor}) by multiplying the kernel density estimates of the two posterior distributions and integrating them. Also we compute $\Rlu$ using the time delay estimates between the event pairs. Figure~\ref{fig:Bul_Bdeltat_scatter_dominik_p2_5yr} shows a scatter plot of the two Bayes factors $\Blu$ and $\Rlu$ estimated from one set of simulated lensed and unlensed events. 

\begin{figure}[hbt] \begin{center}
		\includegraphics[width=3.6in]{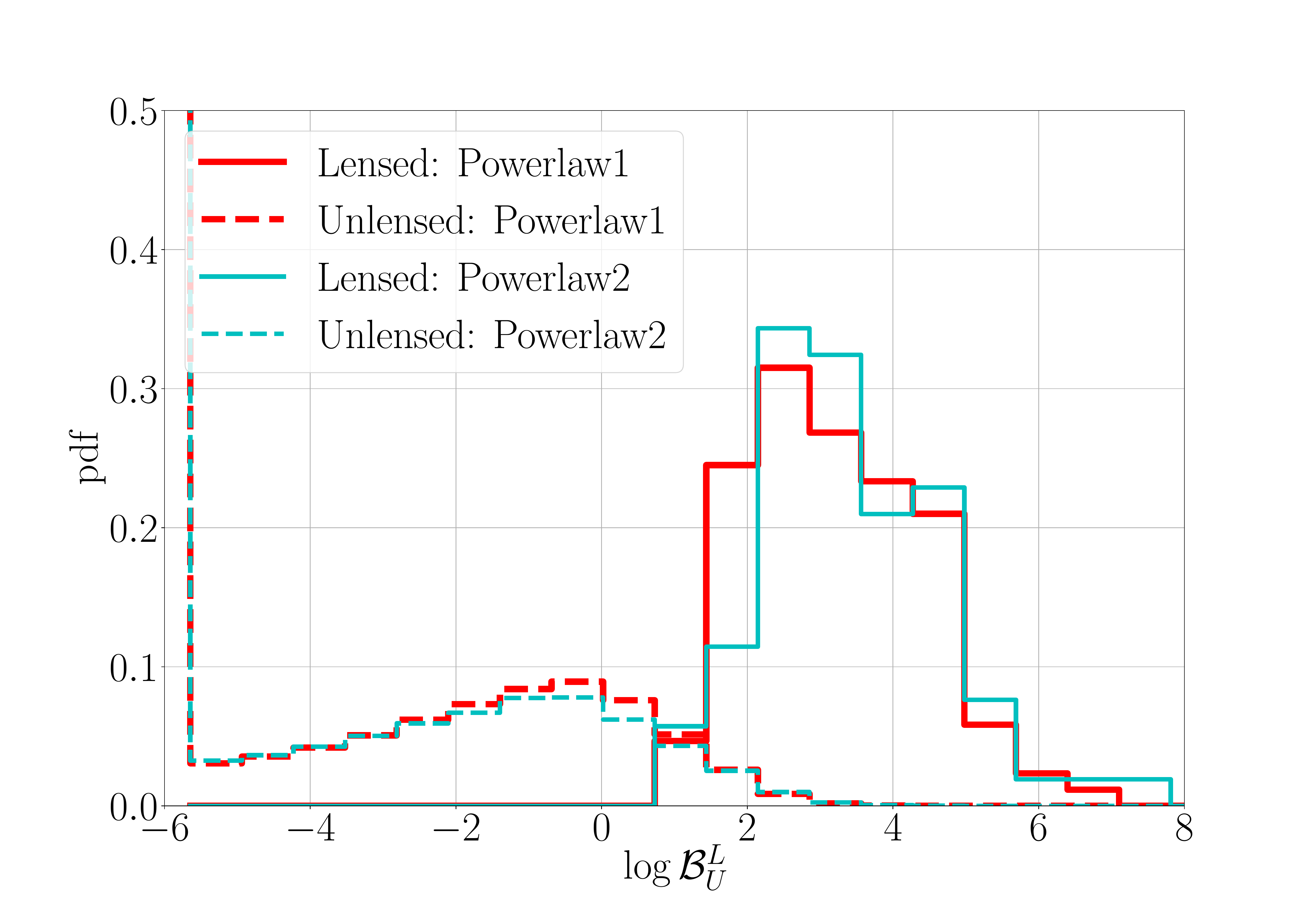}
		\caption{Distribution of the $\log_{10}$ Bayes factor $\Blu$ computed from the \emph{unlensed}  and \emph{lensed}  simulations  with component masses sampled from power law 1 and power law 2.  The Bayes factors are computed using the marginalized posteriors on parameter set $(\mz_1, \mz_2, \cos \alpha, \delta, \chi_1,\chi_2)$. It can be seen that the distributions are not strongly dependent on the specific mass distribution chosen.} 
		\label{fig:Bayes_factor_hist_dominik_p1}
\end{center} \end{figure}
	
\begin{figure*}[tbh] \begin{center}
			\includegraphics[width=3.45in]{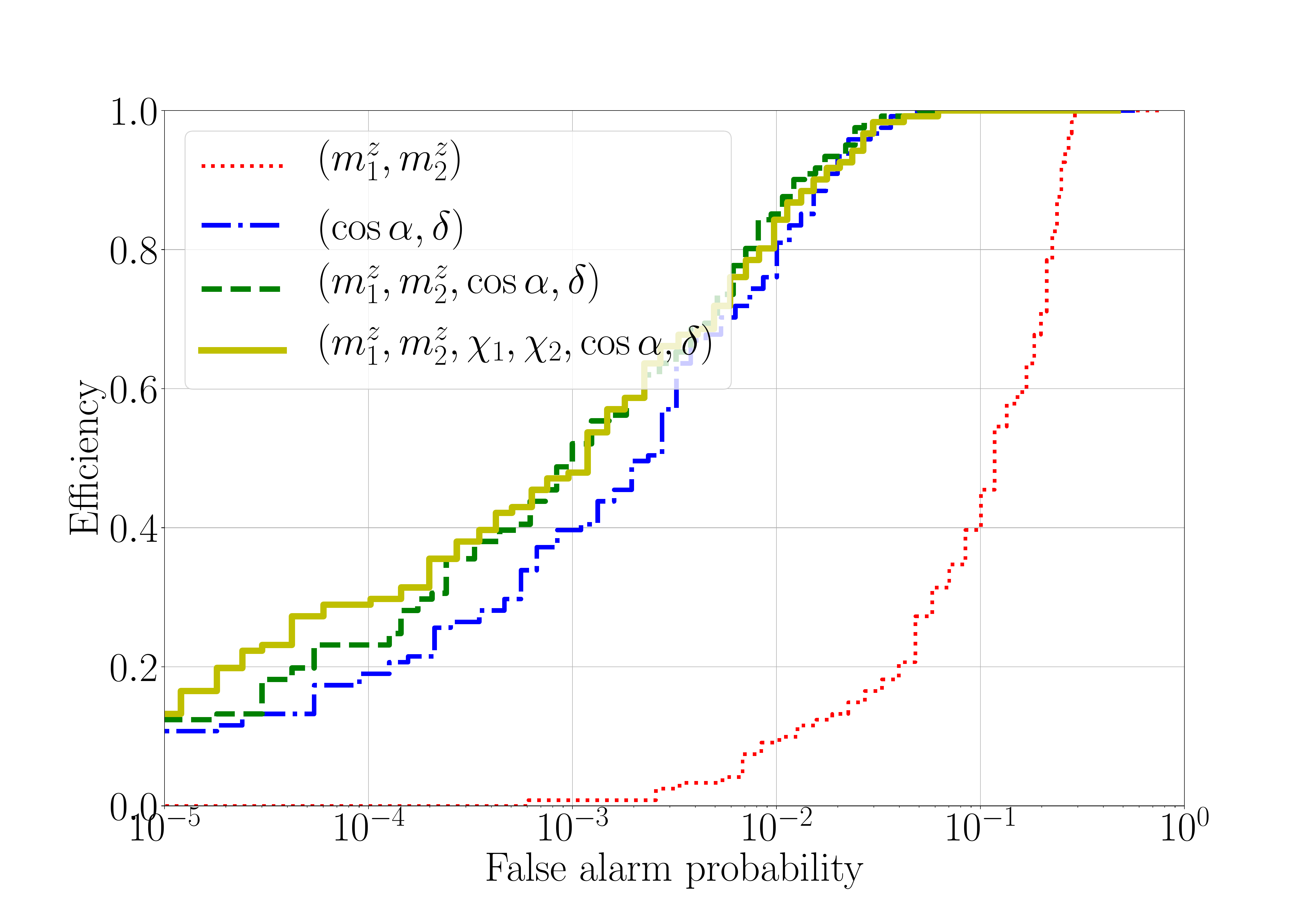}
			\includegraphics[width=3.45in]{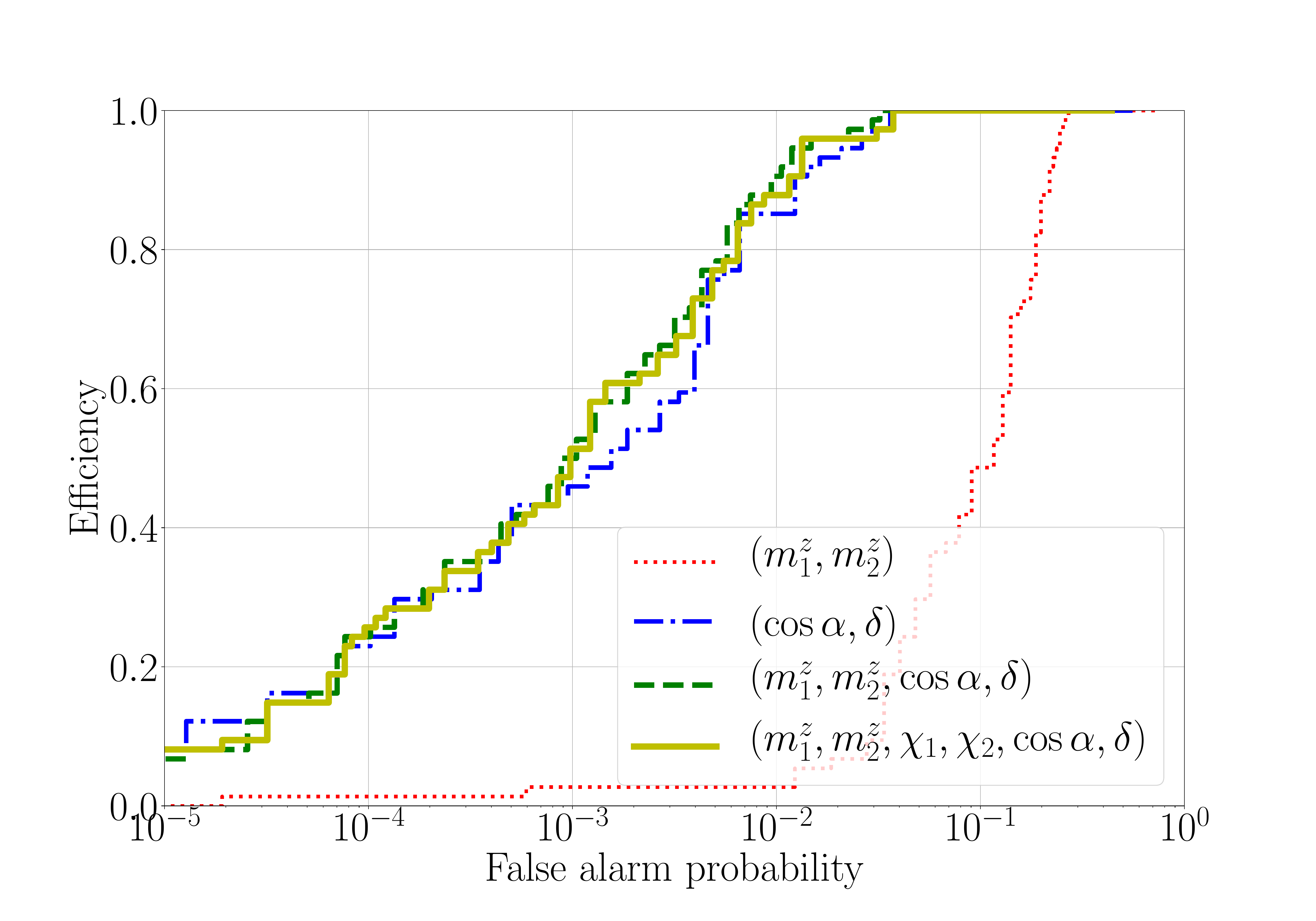}
			\caption{Receiver operating characteristic curves for the  Bayes factor statistic $\Blu$ computed using the marginalized posteriors on parameter sets $(\mz_1,mz_2)$, $(\cos \alpha, \delta)$, $(\mz_1, \mz_2, \cos \alpha, \delta)$  and $(\mz_1, \mz_2,\chi_1,\chi_2 \cos \alpha, \delta)$ respectively  with component masses sampled from power law 1 (left panel) and power law 2 (right panel). We observe that the performance of  the  statistic improves with  with number of parameters. $\Blu$ computed with $(\mz_1, \mz_2,\chi_1,\chi_2 \cos \alpha, \delta)$ posteriors  identifies $\sim 10 - 15\%$ of the lensed event pairs with a false alarm probability of $10^{-5}$.} 
			\label{fig:Bayes_factor_roc_dominik}
		\end{center} 
\end{figure*}

\begin{figure*}[tbh] \begin{center}
		\includegraphics[width=3.45in]{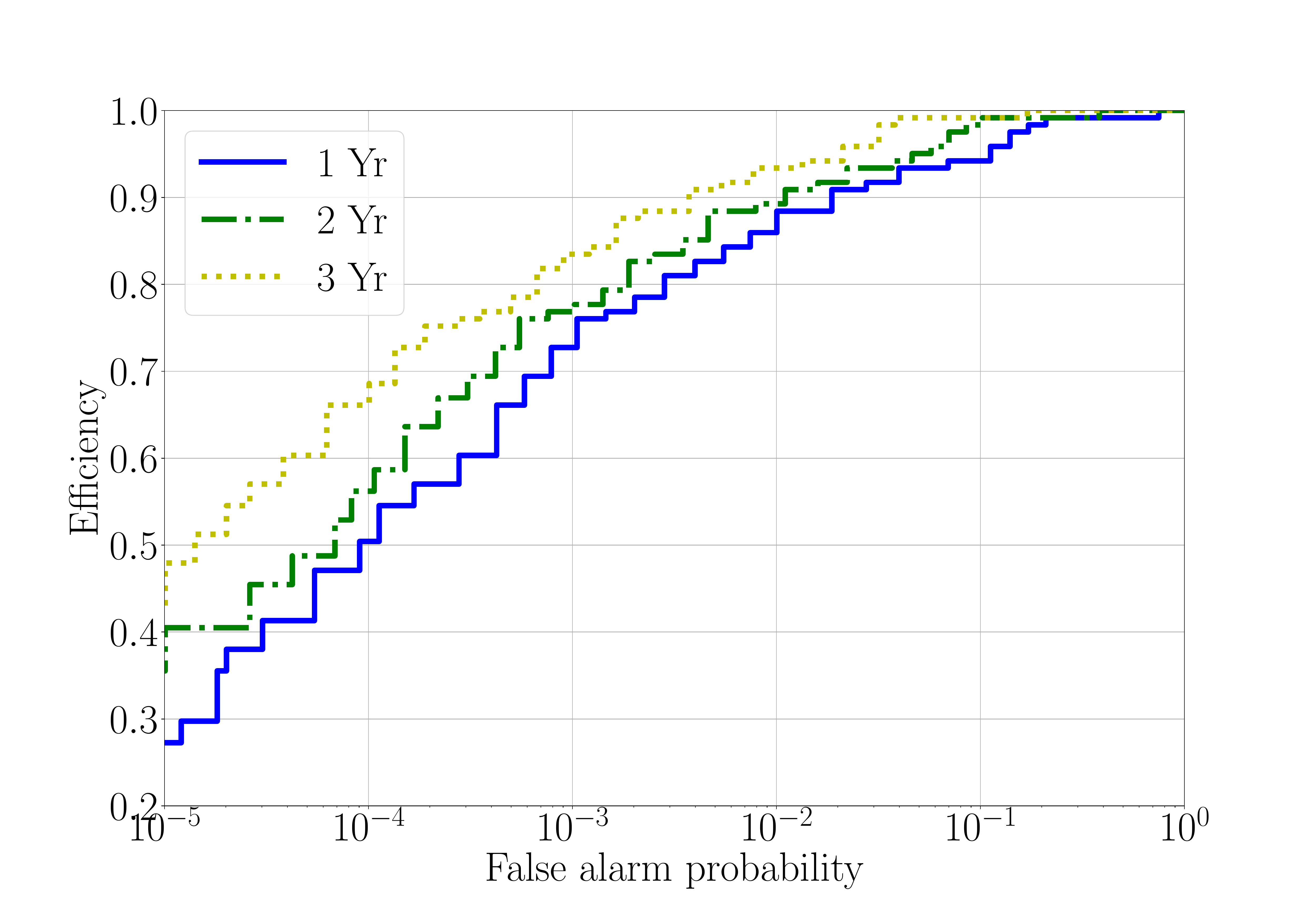}
		\includegraphics[width=3.45in]{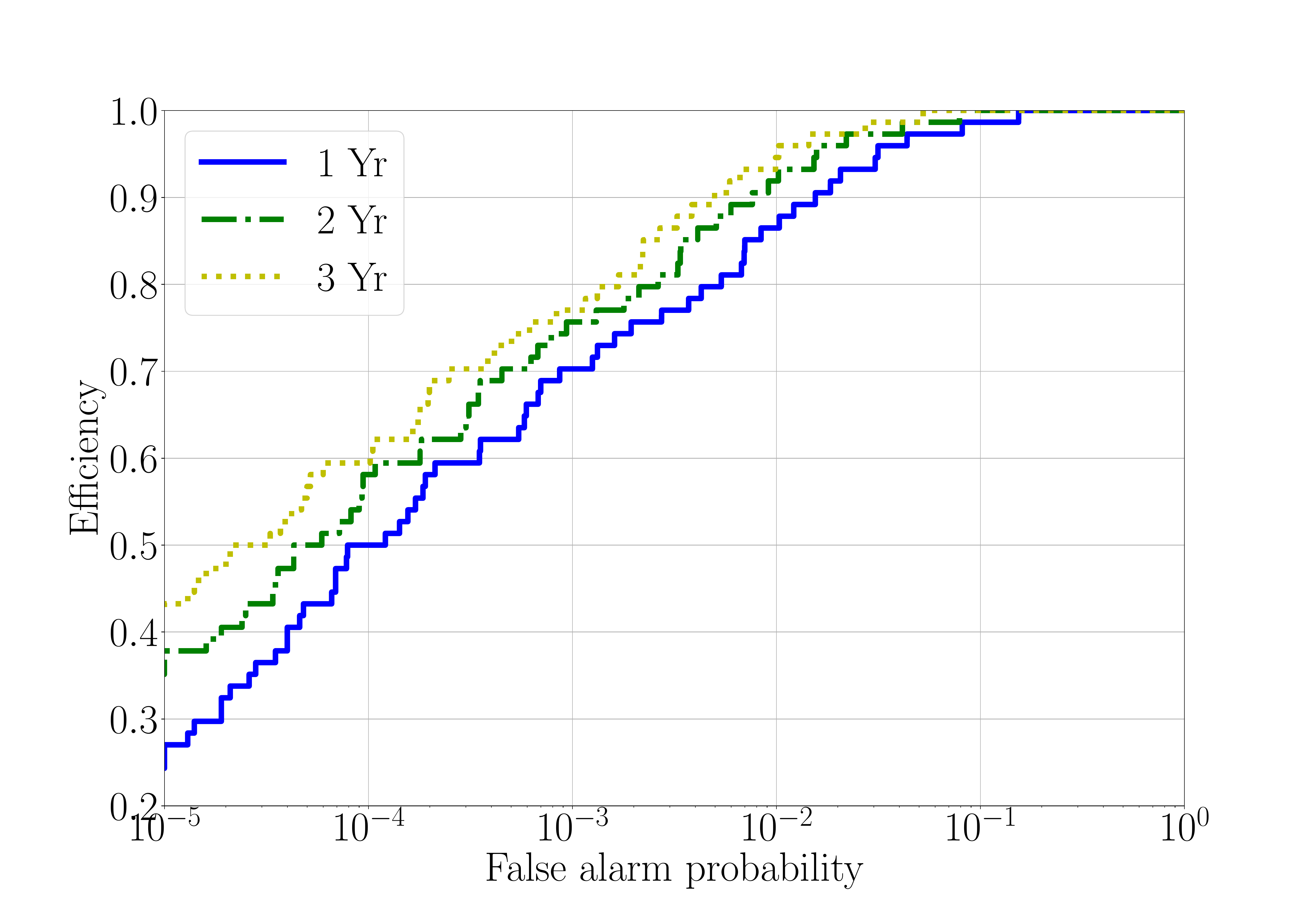}
		\caption{Receiver operating characteristic curves for the  $\Rlu$ statistic computed assuming a rate of 10 unlensed events per month and component masses sampled from power law 1 (left panel) and power law 2 (right panel). Three curves in each panel represent the ROC plots for $\Rlu$ computed assuming 1, 2 and 3 years as the observation time.} 
		\label{fig:Tlag_Bayes_factor_roc_dominik}
\end{center} \end{figure*}

Figure \ref{fig:Bayes_factor_hist_dominik_p1}  shows the distributions of $\log \Blu$ for lensed and unlensed event pairs computed from the posteriors of $\{m^z_1, m^z_2, \cos \alpha, \delta, \chi_1, \chi_2\}$. Indeed, there is a small probability that two independent event pairs could have parameters that  appear mutually consistent (accidentally) and produce a large value for $\Blu$ (``false alarm''). Similarly, the statistic $\Blu$ computed for a truly lensed pair could sometimes attain small values (e.g., due to fluctuations in the detector noise), and reduce the efficiency for detecting truly lensed events. This causes the distributions of the Bayes factor computed from lensed and unlensed events to overlap; a good discriminator should minimize this overlap. Figure~\ref{fig:Bayes_factor_roc_dominik} shows this efficiency for correctly identifying truly lensed events, as a function of the false alarm probability (probability of wrongly identifying unlensed events as lensed events). We show such receiver operating characteristic (ROC) plots for $\Blu$ computed using different sets of parameters. We see that the discriminating efficiency of the Bayes factor increases when we add more signal parameters while computing the statistic. The source sky location parameters $(\cos \alpha, \delta)$ are the ones that most significantly improve the performance.  However, considering the fact that the expected rate of lensed events is very small ($< 1\%$ of all events), the ROC curves indicate that $\Blu$, by itself, is not a very efficient statistic for identifying lensed events. The detection efficiency of $\Blu$ computed using 6 dimensional posteriors is $\sim 10 - 15\%$ for a false alarm probability of $10^{-5}$.

Similarly, in Fig.~\ref{fig:Tlag_Bayes_factor_roc_dominik} we plot the ROC curves for the time-delay Bayes factor $\Rlu$ computed for the same simulated injected events with an average rate of 10 events per month as the binary black hole detection rate. The three curves represent the ROC plots for $\Rlu$ computed assuming 1, 2 and 3 years of observation time. The efficiency of $\Rlu$ increases with the  total length of  the observation time included in the analysis. This is because the distribution of the time delay between unlensed event pairs becomes more and more skewed  towards high values as the observation time increases (see Fig.~\ref{fig:Time_delay_hist}).  The performance of $\Rlu$ is better than that of $\Blu$, with an efficiency of $\sim 45 - 50\%$ corresponding to a false alarm probability of $10^{-5}$ for an observation time of 3 years.

\begin{figure}[tbh] \begin{center}
		\includegraphics[width=3.4in]{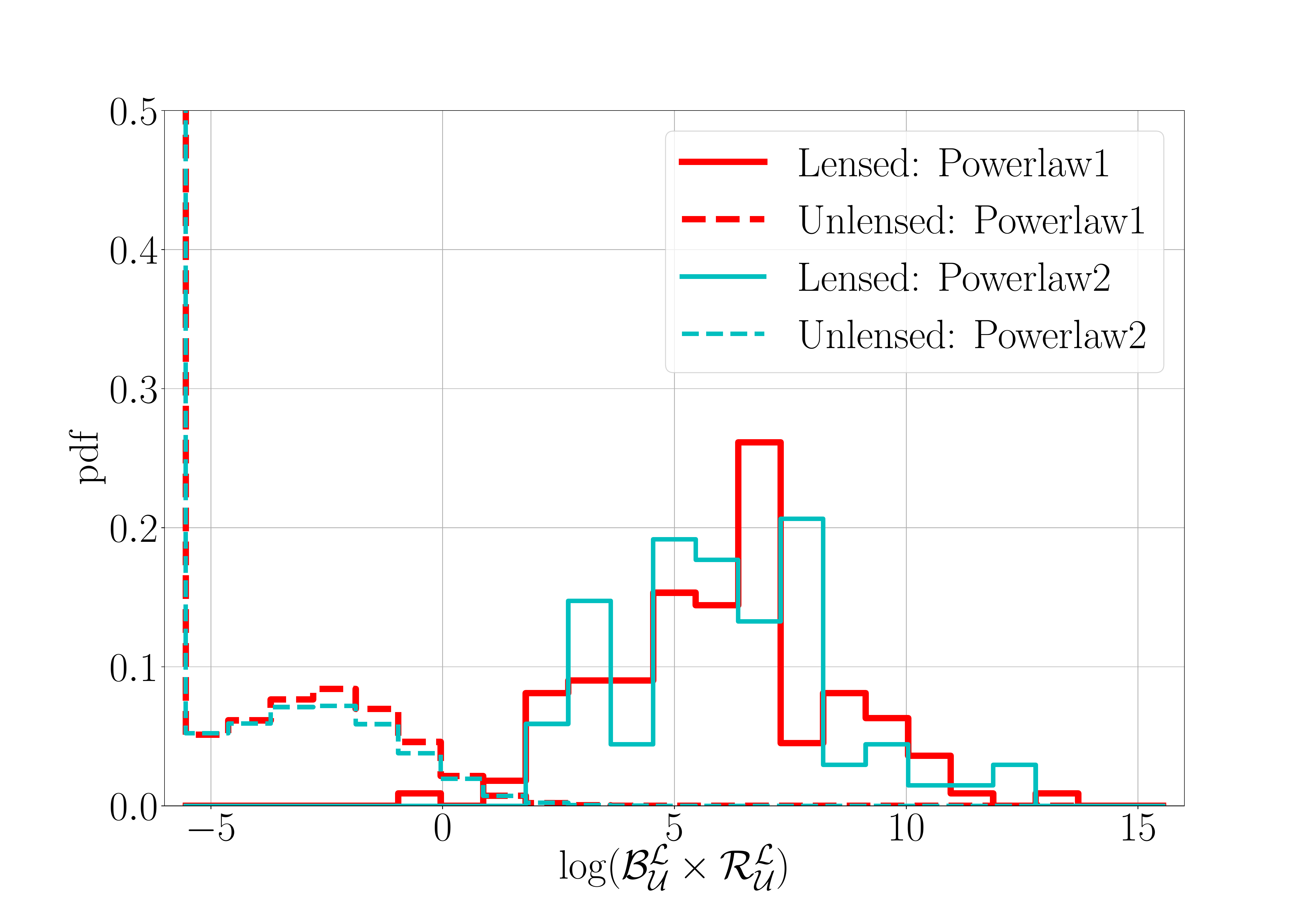}
		\caption{Distribution of the logarithm of the combined  Bayes factor  computed from the \emph{unlensed} and \emph{lensed} simulations  with component masses sampled from power law 1 (solid) and power law 2 (dashed). The Bayes factors are computed using the marginalized posteriors on parameter set $(\mz_1, \mz_2, \cos \alpha, \delta, \chi_1,\chi_2)$.  We use one year of unlensed events for the simulation.} 
		\label{fig:CombinedBayes_factor_hist_dominik_p1_1yr}
\end{center} \end{figure}

\begin{figure}[tbh] \begin{center}
		\includegraphics[width=3.4in]{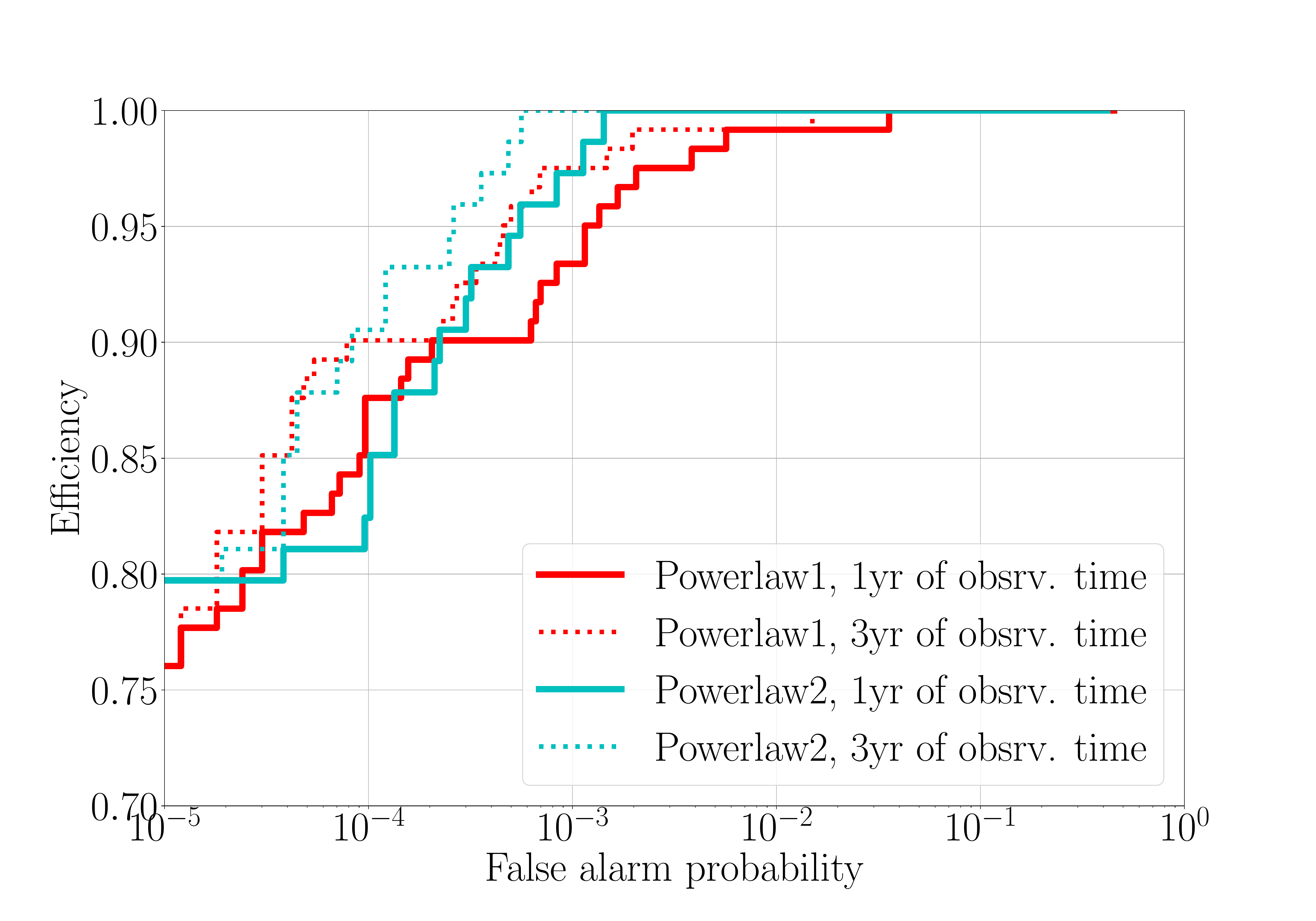}
		\caption{Receiver operating characteristic curves for the combined  Bayes factor statistic computed using the marginalized posteriors on parameter set $(\mz_1, \mz_2,\chi_1,\chi_2 \cos \alpha, \delta)$ with component masses sampled from $\Pr_1(m_1,m_2)$ (solid) and $\Pr_2(m_1,m_2)$ (bashed). The statistic is able to correctly identify $\sim 80\%$ of the lensed events with a false alarm probability of $10^{-5}$.}
		\label{fig:CombinedBayes_factor_roc_dominik_1yr}
\end{center} \end{figure}

As one can see in the scatter plot of $\Blu$ and $\Rlu$ of lensed/unlensed events pairs in Fig.~\ref{fig:Bul_Bdeltat_scatter_dominik_p2_5yr}, applying individual thresholds on $\Blu$ (vertical) and $\Rlu$ (horizontal) are less effective in separating lensed pairs (red triangles) from unlensed pairs (blue stars). However, a combined threshold can improve the discriminatory power. Therefore, as described in Sec. \ref{bayes_facotr_formulation}, we combine $\Blu$ with $\Rlu$ and define their product as a new statistic. Figure \ref{fig:CombinedBayes_factor_hist_dominik_p1_1yr} shows the  distributions of this combined statistic for lensed and unlensed event pairs with one year of observation time. Figure~\ref{fig:CombinedBayes_factor_roc_dominik_1yr} shows the ROC plots for this combined statistic computed assuming 1 and 3 years of observations time. The results clearly demonstrate that the combined statistic has a significantly higher detection efficiency when compared to $\Blu$ and $\Rlu$. For a false alarm probability of $10^{-5}$, the product statistic (computed using the six dimensional posteriors) identifies $\sim 80\% $ of the lensed event pairs. 


\section{Summary and future work}
\label{sec:conclusions}

In this paper we propose a method for statistically identifying multiple images of strongly lensed binary black hole merger events from a population of GW detections by the LIGO-Virgo network. Recent estimates show that Advanced LIGO and Virgo, when they reach their design sensitivities, will detect several binary black hole mergers per year that are strongly lensed by intervening galaxies~\cite{Ng:2017yiu}. We will be able to observe multiple images of such GW signals, which are separated by time scales of minutes to weeks. In the case of GW signals from stellar mass black-hole binaries lensed by galaxies (for which $\lambda_\mathrm{GW} \ll 2 G M_\mathrm{lens}/c^2$), the lensing will result in a magnification/de-magnification of the GW polarizations without affecting their frequency profile. Hence, the parameters of the binary that determine the frequency evolution of the signal (such as the redshifted masses and spins), which we  extract from multiple images, will be mutually consistent~\footnote{Note, however, that the luminosity distance that we extract using the parameter estimation using standard (unlensed) templates will be biased, due to the unknown magnification in the signal. Hence the inferred redshift and intrinsic masses will also be biased~\cite{2017PhRvD..95d4011D}}. In addition, since the deflection angle is small compared to the typical source-localization accuracies, the sky-location of multiple images will also be the same. In order to determine whether a pair of binary black hole signals are lensed images of the same merger, we check the consistency of extracted parameters (except the luminosity distance) from the two signals. To be precise, we computed the odds ratio between two hypotheses 1) that they are the lensed images of the same merger event, 2) that they are two unrelated events. This odds ratio can be written in terms of the overlap of the posterior distributions of the extracted parameters from the two events, inversely weighted by the prior [see Eq.~\eqref{eq:lensing_bayes_factor}]. In addition, we make use of the fact that the distribution of the time delays between a pair of lensed events will be different from that between a pair of random uncorrelated events (see Fig.~\ref{fig:Time_delay_hist}). This allows us to define another odds ratio between the two hypotheses based on the observed time delay between a pair of events [see Eq.~\eqref{eq:bayesfactor_lensing_timedel}]. We combine these two different odds ratios to form a more sensitive discriminator between lensed and unlensed events. 

We test the efficiency of the proposed statistic by simulating binary black hole merger events in the LIGO-Virgo network with design sensitivity. The simulations shows that the pipeline can distinguish images $\sim 80\%$ of strongly lensed merger events from unlensed events with a false alarm probability of $10^{-5}$ for  three years of observation time. 

There are possible ways of improving the discriminatory power of this statistic: one is by increasing the number of parameters that are used to test the consistency between estimated parameters of the two events (e.g., inclination angle, spin orientations, etc., if they are well measured). Secondly, one can use the property discovered by~\cite{Dai:2017huk} that waveforms of different images are related by specific phase shifts. Thirdly, one could explore the possibility of using priors on the magnification ratios of multiple images (or the ratios of the SNRs of multiple images) in a way similar to the way we used the priors on time delays between multiple events to distinguish between lensed and unlensed pairs. We leave these as future work. 

\acknowledgments 
The authors thank Tjonnie Li for the careful reading of the manuscript and his useful comments. This research was supported by the Indo-US Centre for the Exploration of Extreme Gravity funded by the Indo-US Science and Technology Forum (IUSSTF/JC-029/2016).  TV acknowledges support from the Schmidt Fellowship, and the W.M. Keck Foundation Fund. PA's research was supported by the Science and Engineering Research Board, India through a Ramanujan Fellowship, by the Max Planck Society through a Max Planck Partner Group at ICTS-TIFR, and by the Canadian Institute for Advanced Research through the CIFAR Azrieli Global Scholars program. SK acknowledges support from national post doctoral fellowship (PDF/2016/001294) by Scientific and Engineering Research Board, Govt. of India. Computations were performed at the ICTS cluster Alice. 

\appendix

\section{Generating samples of strongly lensed and multiply imaged binary mergers}
\label{lens_simulation}
In this section, we outline our method for generating samples of strongly lensed and multiply imaged binary merger events. We will use results for strong lensing probabilities that have been derived earlier (see~e.g., \cite{1994A&A...284..285K,2015ApJ...811...20C}). Given below is a brief summary of our method and assumptions:
\begin{enumerate}
	\item Given a source redshift, the bulk of the magnification probability describes cases with a single lensed image~\cite{2011ApJ...742...15T,2017PhRvD..95d4011D}. We are interested in multiply imaged mergers, so we do not need to accurately model the cases with single images. 
	\item Multiple images dominantly arise due to galaxy lenses \cite{1991MNRAS.253...99F}. We model individual strong lenses as isothermal ellipses with non-zero ellipticity. 

	\item Singular isothermal ellipsoid (SIE) lens models have a surface mass density that diverges at the center. These lenses produce either two or four images~\cite{1994A&A...284..285K}.
	
	\item The lens model has two parameters: velocity dispersion $\sigma$ and axis-ratio $q$. We  generate these parameters with distributions taken from the SDSS galaxy population  \cite{2015ApJ...811...20C}. The axis-ratio does not dramatically change the strong lensing cross section, so we can estimate overall rates in the manner of Ref.~\cite{2011Natur.469..181W}.
	
\end{enumerate}

\subsection{Probability of multiple imaging}
\label{sec:calc}

Given the assumptions that are outlined above, the multiple imaging optical depth $\tau(\zs)$ to a 
given source redshift $\zs$ is~\cite{2011Natur.469..181W}:
\begin{align}
	\tau(\zs) & = \int_0^{\zs} \frac{d\tau}{dz_{l}} d\zl,
\end{align}
where the differential optical depth per unit lens redshift $\zl$ is
\begin{align}
	\label{eq:diffod}
	\frac{d\tau}{d\zl} & = \int d\sigma \, n(\zl) \frac{dp}{d\sigma} (1+\zl)^3 \frac{cdt}{d\zl} \pi \Dl(\zl)^2 \theta^2(\sigma, \zl, \zs).
\end{align}
Here, $\sigma$ is the lens' velocity dispersion, $n(\zl)$ is the comoving number  density of lenses, $dp/d\sigma$ is the PDF of the velocity dispersion $\sigma$, $\Dl$ is the angular diameter distance to the lens, and $\theta$ is the angular Einstein radius of a singular isothermal sphere (SIS) lens. We assume a constant number density and an unchanging PDF of the velocity dispersion, which are reasonable for galaxy lenses at relatively low redshifts~\cite{2011ApJ...737L..31B}. 

Let us start with the parameters for the population of early-type galaxies from Ref.~\cite{2007ApJ...658..884C}: the number density $n = 8 \times 10^{-3} h^3 {\rm Mpc}^{-3}$, and the distribution of velocity dispersion (VDF) is

\begin{align}
	\label{eq:sigpdf}
	\frac{dp}{d\sigma} & = \left( \frac{\sigma}{\sigma_*} \right)^\alpha 
	\exp{\left[ -\left( \frac{\sigma}{\sigma_*} \right)^\beta \right]} \frac{\beta}{\Gamma(\alpha/\beta)}
	\frac{1}{\sigma},
\end{align}
where $\alpha = 2.32$, $\beta = 2.67$, and $\sigma_* = 161 \, {\rm km \, s^{-1}}$. Substituting 
the Einstein radius for a SIS $\theta = 4\pi (\sigma^2/c^2) \Dls/\Ds$ in Eq.~\eqref{eq:diffod}, 
we get
\begin{align}
\frac{d\tau}{d\zl} & = 16\pi^3 (1+\zl)^2  \frac{c \, n}{H(\zl)} \left( \frac{\Dl \Dls}{\Ds} \right)^2 \left( \frac{\sigma_*}{c} \right)^4
\frac{\Gamma([4+\alpha]/\beta)}{\Gamma(\alpha/\beta)}.
\end{align}
The total multiple-imaging optical depth is
\begin{align}
\begin{split}
\tau(\zs) & = 16\pi^3 c n \left( \frac{\sigma_*}{c} \right)^4 \frac{\Gamma(\frac{4+\alpha}{\beta})}{\Gamma(\alpha/\beta)} \int_0^{\zs} d\zl (1+\zl)^2 \frac{1}{H(\zl)} \left( \frac{\Dl \Dls}{\Ds} \right)^2 \\
& = 16\pi^3 n \left( \frac{\sigma_*}{c} \right)^4 \frac{\Gamma(\frac{4+\alpha}{\beta})}{\Gamma(\alpha/\beta)} \int_0^{\Dsc} d\Dlc \, {\Dlc}^2 \left( 1 - \frac{\Dlc}{\Dsc} \right)^2 \label{eq:tausint}  
\end{split}
\\
\begin{split}
& = 16 \pi^3 \, \left( \frac{\sigma_*}{c} \right)^4 \frac{\Gamma(\frac{4+\alpha}{\beta})}{\Gamma(\alpha/\beta)} \frac{n \, {\Dsc}^3}{30} \\ 
& = 4.17 \times 10^{-6} \left( \frac{\Dsc}{\rm Gpc} \right)^3. 
\label{eq:lensing_prob}
\end{split}
\end{align}
In the last line, we have written the result in terms of the comoving distance $\Dc(z) = \int_{0}^{z} dz^\prime \, c/H(z^\prime)$, and used $\Dlc$ and $\Dsc$ to denote $\Dc(\zl)$ and $\Dc(\zs)$, respectively. For the simulations this paper, we use the following values for the cosmological parameters in the $\Lambda$CDM model: $H_0 = 70 \text{Km\,s}^{-1}\text{Mpc}^{-1}$ and $\Omega_\Lambda = 0.7$.

Ref.~\cite{Ng:2017yiu} use a similar scaling as in Eq.~\eqref{eq:lensing_prob} for the strong lensing optical depth. However, their normalization (as derived in Ref.~\cite{1991MNRAS.253...99F}) is larger by a factor of $6.3$. The difference arises because the number density and VDFs provided in Ref.~\cite{2007ApJ...658..884C} are fits to the SDSS population of early-type galaxies, which dominate the high velocity-dispersion end (and can be dominantly selected for in strong lensing surveys). Ref.~\cite{2010MNRAS.404.2087B} provide the number densities and VDFs for the entire galaxy population, and obtain a similar enhancement in the total characteristic number density $n$ (and even larger characteristic velocity dispersions for early-type galaxies). The selection effects for GW lensing are very different from those for optical surveys (obscuration by the stellar light from the lens galaxy is not an issue), and hence, it is appropriate to use all lens galaxies when forward-modeling the population of lensed sources. However, this difference is immaterial for our study. 

\subsection{Method to generate samples of lensed events}
\label{sec:sampling}

In this section, we outline our method for drawing samples of strongly lensed mergers from a given source distribution.
\begin{figure*}[ht] \begin{center}
		\includegraphics[width=3.45in]{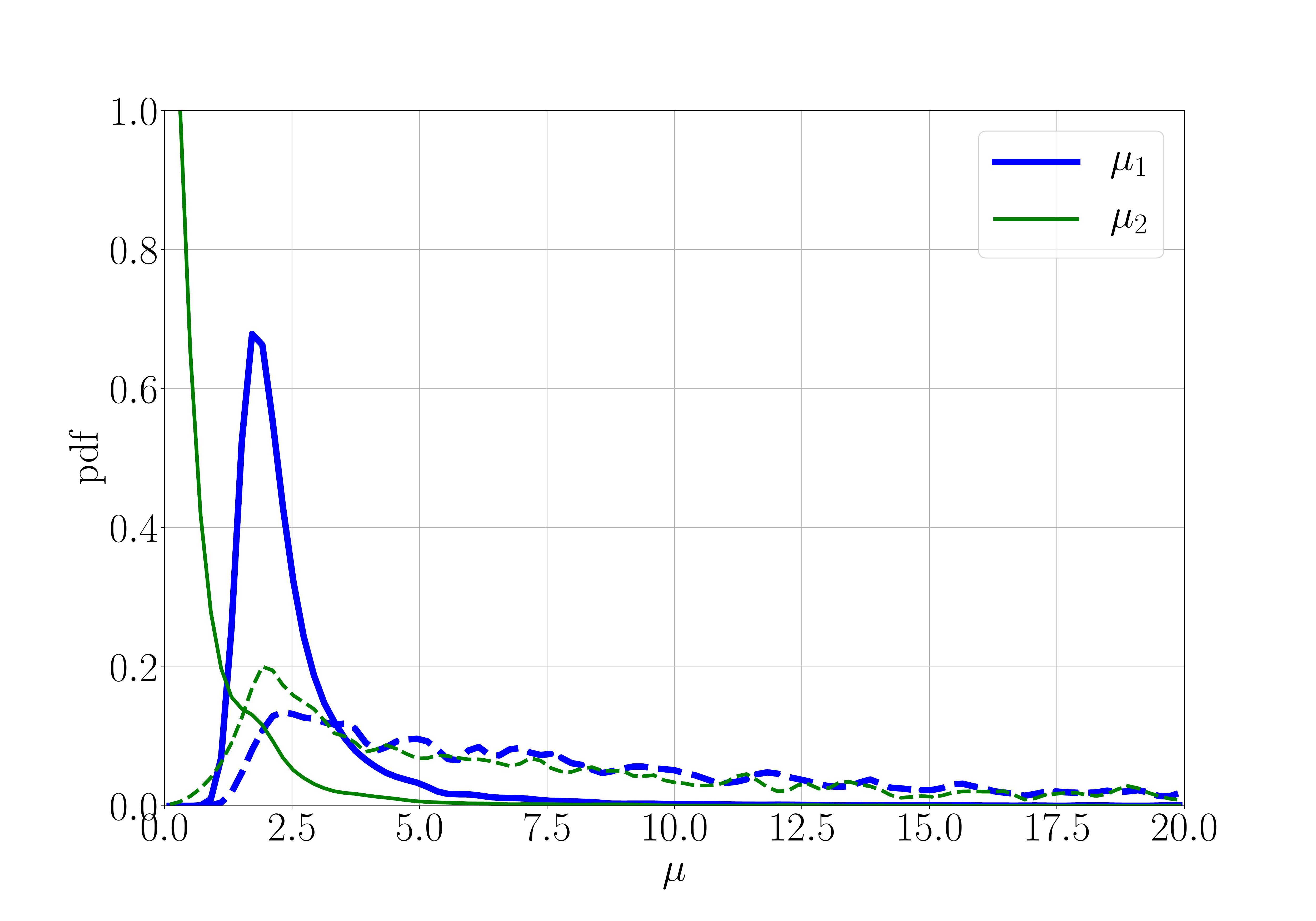}
		\includegraphics[width=3.45in]{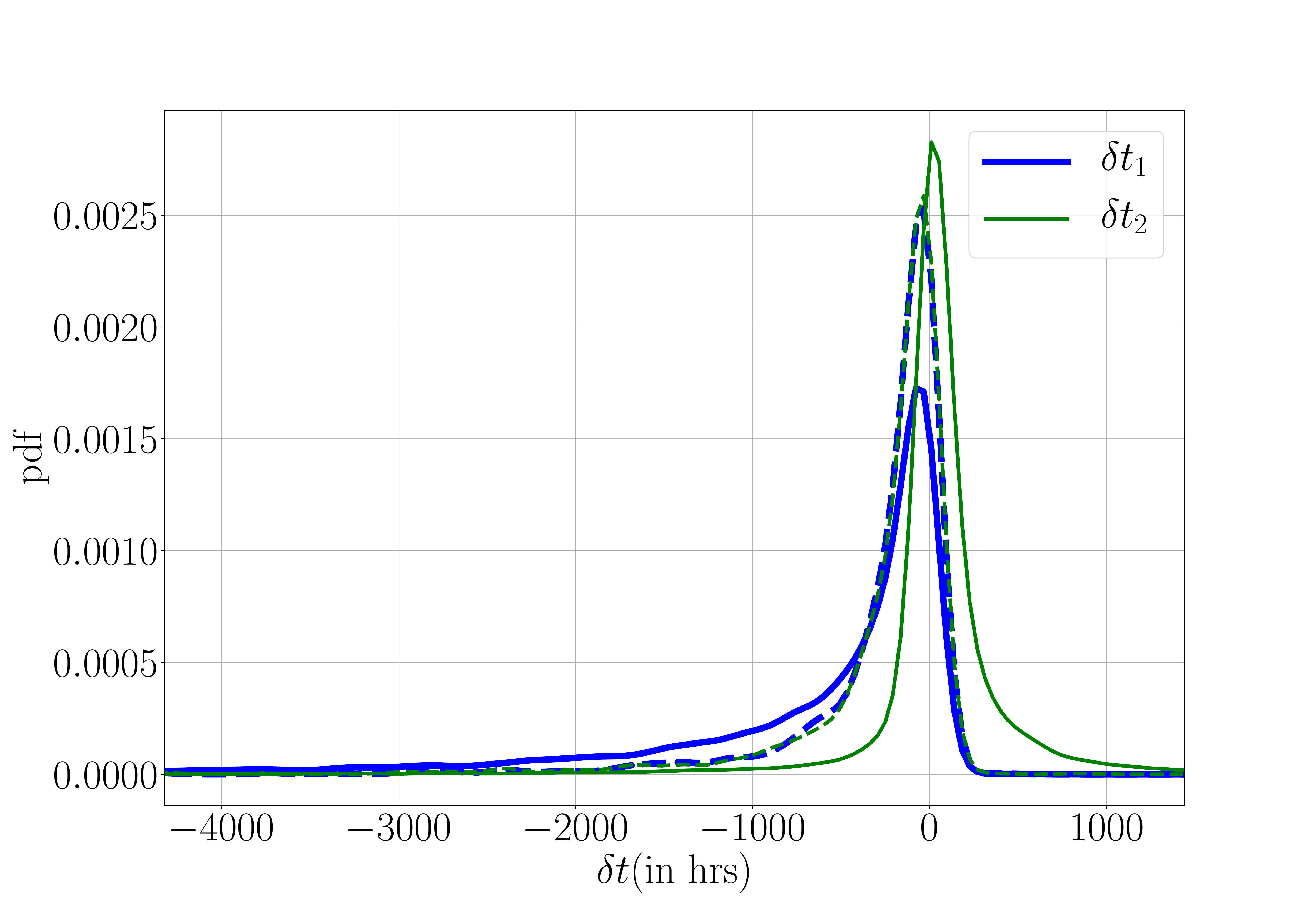}
		\caption{Distributions of the magnifications $\mu_1, \mu_2$  (left) and the arrival times $\delta t_1, \delta t_2$ relative to unlensed arrival time (right) of  the two dominant images   for simulated events (See Eqs. \ref{eq:mu} ans \ref{eq:deltat} ). Solid (dashed) traces show distributions before (after) applying the detection threshold SNR $\geq 8$. The component masses of the simulated events are sampled from power law 1 distribution.}
		\label{fig:mu_deltat_dist}
	\end{center} 
\end{figure*}
\begin{enumerate}
	\item \emph{Pick a source:} We start with a merger whose intrinsic parameters (total mass $M = m_1 + m_2$, 
	symmetric mass ratio $\eta = m_1 m_2/M^2$, and dimensionless spins $\chi_1$ 
	and $\chi_2$) are drawn from given distributions. 
	In addition, we randomly draw the angles ($\iota, \psi$)
	associated with the binary's plane so that its orbital angular momentum is distributed 
	uniformly over the sphere, and randomly draw its position ($\cos \alpha, \delta$) so that the binaries are 
  uniformly distributed in the sky. The redshift $\zs$ is distributed as given in  \cite{Dominik:2013tma}	(see, Fig.~\ref{fig:Dominik_z_powerlaw_m_hist}). See, Sec.~\ref{sec:astro_sim} for more details. 
	\item \emph{Accept/reject according to the multiple imaging probability}: Given the source redshift $\zs$, we read 
	off the multiple--imaging probability $\tau(\zs)$ from (the enhanced version of) Eq.~\eqref{eq:lensing_prob}. If $\tau(\zs)$
	is larger than a random number uniformly distributed between 0 and 1, we proceed to step 3. If not, we discard this source. 
	\item \emph{Draw the lens redshift:} If the merger survives step 2, we draw a sample $r$ from 
	the PDF
	\begin{align}
	p(x) = 30 \, x^2 (1 - x)^2, \qquad 0<x<1.
	\end{align}
	and compute a sample lens comoving distance using $D^c(\zl) = r \, D^c(\zs)$; we obtain the 
	lens redshift $\zl$ by inverting $D^c(\zl)$. Using 
	Eq.~\eqref{eq:tausint}, we see that if a source at $\zs$ is multiply imaged, this procedure yields lens redshifts with the right posterior distribution. 
\item \emph{Draw the lens parameters:} We use the fits for the distribution of the lens parameters 
from Ref.~\cite{2015ApJ...811...20C}. We draw a parameter $a$ from a generalized Gamma distribution
\begin{align}
p(x) & = x^{\alpha-1} \exp{\left( -x^\beta \right)} \ \frac{\beta}{\Gamma(\alpha/\beta)},
\end{align}
where $\alpha = 2.32, \beta = 2.67$, and set $\sigma = 161 \, {\rm km \, s^{-1}} \times a$. We next sample the distribution of the axis ratio of the lens.
Given the above sample of $a$, we repeatedly draw parameter $b$ from a Rayleigh distribution
\begin{equation}
g(x) = \frac{x}{s^2} \exp{\left( -\frac{x^2}{2s^2} \right)}, \qquad 0<x<\infty, 
\end{equation}
where,
\begin{equation}
s = 0.38 + 0.09177 \, a,  
\end{equation}
until we get a sample $b < 0.8$. We then set the axis ratio $q = 1 - b$.
\item \emph{Draw a source--plane location:} Given a lens with the above parameters, we then sample the source--plane location of the merger. Since we have already determined that it is multiply imaged, 
we only need to get the right posterior distribution of the source, which is a uniform distribution within 
the cut/caustics of the lens model. A complication is that we cannot analytically 
calculate the intersection of the two and four image regions for small values of the axis ratio. Our approach will be to use the results in 
Ref.~\cite{1994A&A...284..285K}, and draw with repetition. The idea is to repeatedly draw points 
$(y_1, y_2)$ within a certain range, and solve the lens equation (as detailed in Step 6), until we obtain a location with multiple images. 

Given axis ratio $q$, we draw coordinates $y_1$ and $y_2$ from uniform distributions in the following 
ranges:
\begin{align}
y_1 & \in \left(0, \sqrt{ \frac{q}{1-q^2} } \, {\rm arccosh}\left[ \frac{1}{q} \right] \right), \\
y_2 & \in 
\begin{cases}
\left(0, \sqrt{ \frac{q}{1-q^2} } \, {\rm arccos}\left[ q \right] \right), & \mathrm{if}~ q > q_0 \\
\left(0, \sqrt{\frac{1}{q}} - \sqrt{ \frac{q}{1-q^2} } \, {\rm arccos}\left[ q \right] \right), & \mathrm{if}~ q < q_0
\end{cases}
\end{align}
Here $q_0 = 0.3942$ is the numerical solution to the transcendental equation 
$2 q_0 \, {\rm arccos} ~ q_0 = \sqrt{1 - q_0^2}$.
\item \emph{Solve the lens equation:} Given $y_1$, $y_2$, and $q$, we numerically find all roots of the 
one-dimensional equation 
\begin{align}
&\left[ y_1 + \sqrt{ \frac{q}{1-q^2} } {\rm arcsinh}\left( \frac{\sqrt{1 - q^2}}{q} \cos{\phi} \right) \right] \sin{\phi} \,\, - \nonumber \\
&\left[ y_2 + \sqrt{ \frac{q}{1-q^2} } {\rm arcsin}\left( \sqrt{1 - q^2} \sin{\phi} \right) \right] \cos{\phi} = 0
\end{align}
in the interval $[0, 2\pi)$. Assuming that we get solutions $\{ \phi_1, \phi_2, \cdots \}$, we only retain those $\phi_i$ that satisfy the condition
\begin{align}
&\left[ y_1 + \sqrt{ \frac{q}{1-q^2} } {\rm arcsinh}\left( \frac{\sqrt{1 - q^2}}{q} \cos{\phi_i} \right) \right] \cos{\phi_i} \,\, +  \nonumber\\ 
&\quad \left[ y_2 + \sqrt{ \frac{q}{1-q^2} } {\rm arcsin}\left( \sqrt{1 - q^2} \sin{\phi_i} \right) \right] \sin{\phi_i} > 0
\end{align}
If the final list of solutions only contains one element, we go back to Step 5 and repeat until we get a case with a set $\{ \phi_i \}$ with with multiple elements.
\item \emph{Read off image magnifications and time delays:} The deflections are typically small relative to the GW localization uncertainties, so we ignore the differences between image positions on the sky while computing the GW signal. However, we need the positions to calculate the magnifications and time delays from the lens model.

Given the list of solutions $\{ \phi_1, \phi_2, \cdots \}$ from Step 6, and the source position 
$(y_1, y_2)$ for each image, we compute the image positions $(x_{1, i}, x_{2, i})$ as follows:
\begin{align}
x_{1 , i} & = y_1 + \sqrt{ \frac{q}{1 - q^2} } \, {\rm arcsinh}\left( \frac{\sqrt{1 - q^2}}{q} \cos{\phi_i} \right), \\
x_{2 , i} & = y_2 + \sqrt{ \frac{q}{1 - q^2} } \, {\rm arcsin}\left( \sqrt{1 - q^2} \, \sin{\phi_i} \right) 
\end{align}
The magnifications of the images are given by
\begin{align}
\mu_i & = \left({1 - \sqrt{ \frac{q}{x_{1, i}^2 + q^2 x_{2, i}^2} } } \right)^{-1}. \label{eq:mu}
\end{align}
\begin{figure*}[tbh] \begin{center}
		\includegraphics[width=3.45in]{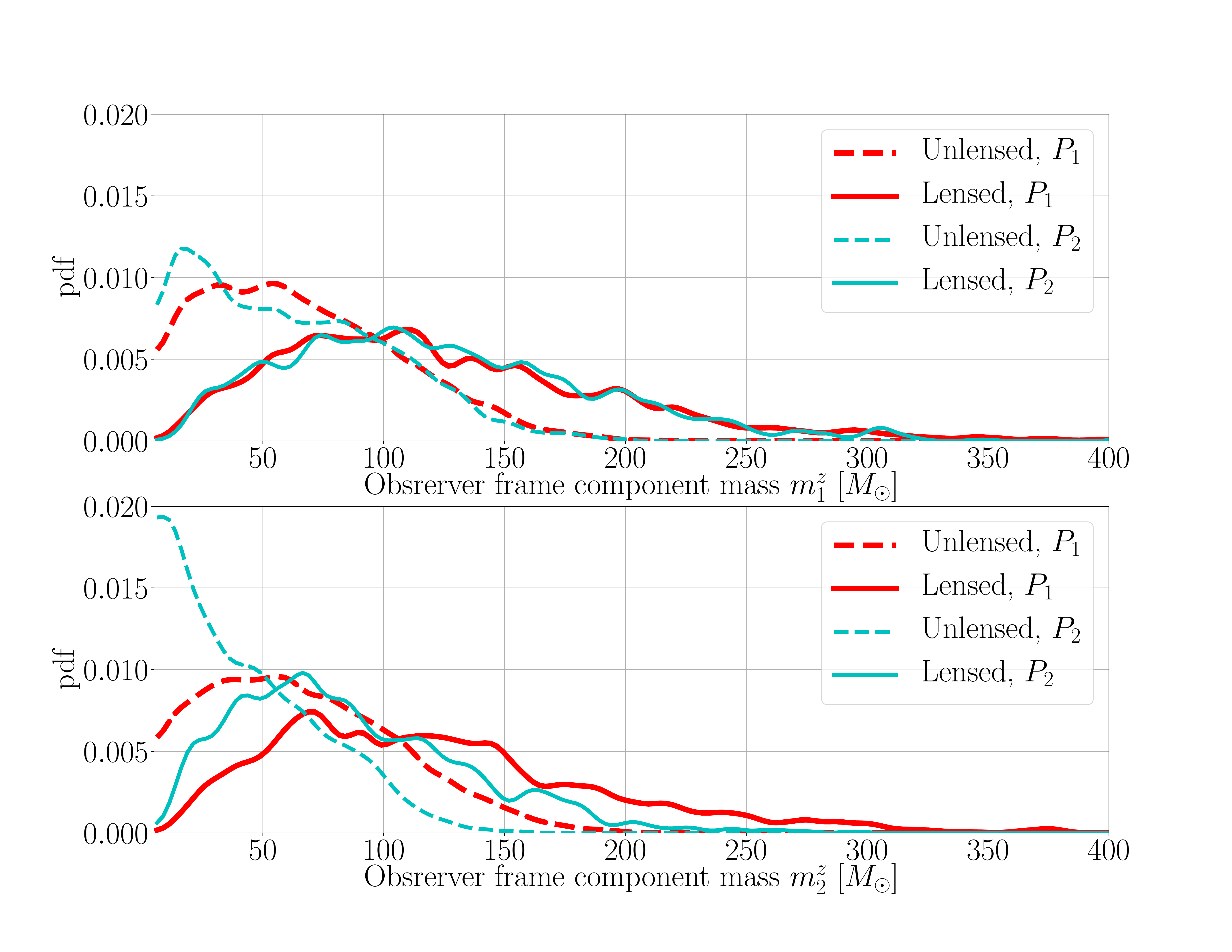}
		\includegraphics[width=3.45in]{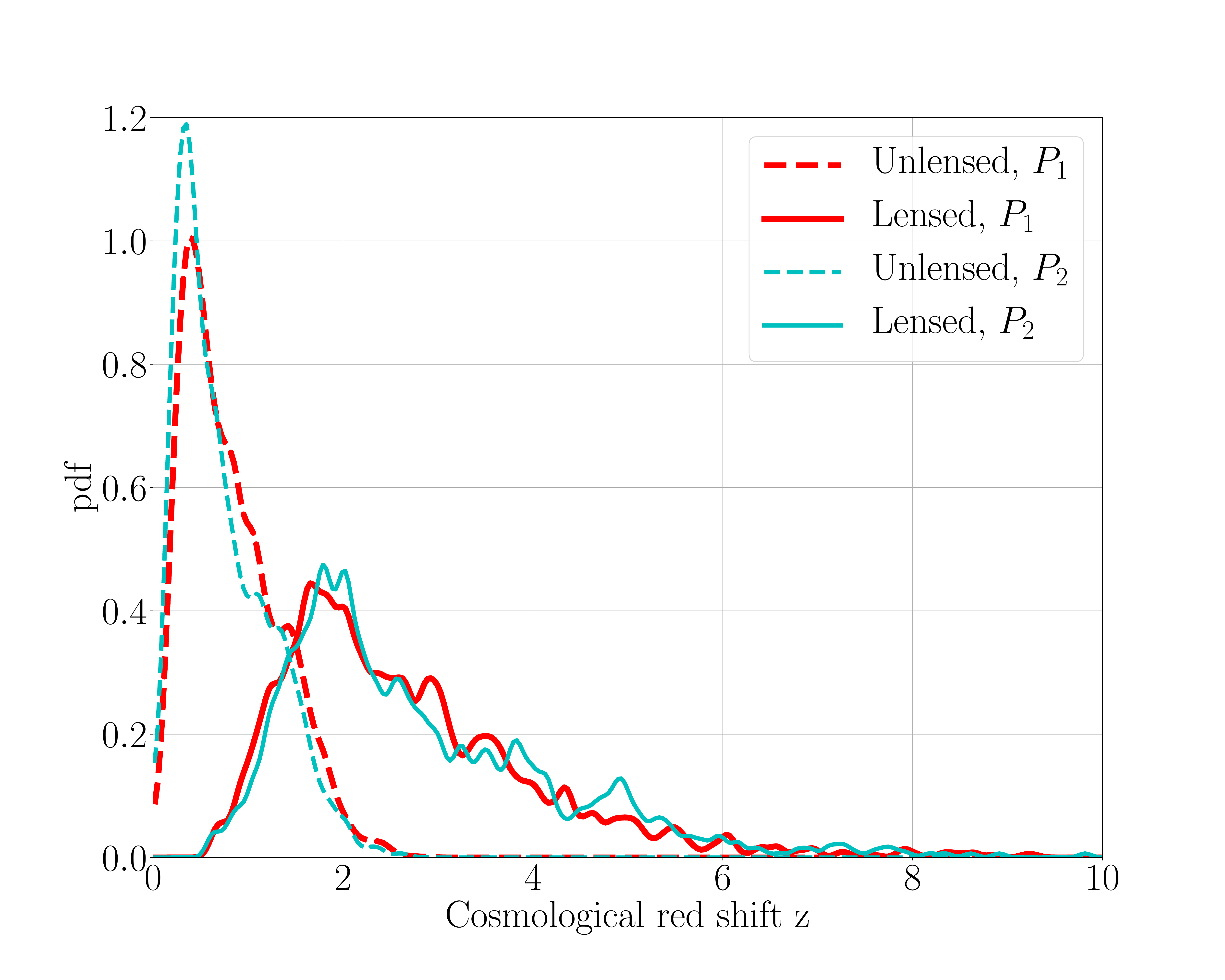}
		\caption{ \emph{Left panel:} The distributions of red shifted component masses $m_1^z$ and $m_2^z$ for unlensed and lensed simulated events producing an SNR $\geq 8$ in the Advanced LIGO-Virgo network. Solid and dashed curved correspond to the source frame mass distributions $P_1$ and $P_2$, respectively.  \emph{Right panel:} The red shift distributions of detectable (SNR $\geq 8$) unlensed and lensed simulated events.}
		\label{fig:injectedredshift_dist}
	\end{center} 
\end{figure*}
The arrival times of the images relative to some common base time), are:
\begin{align}
\delta t_i & = 16\pi^2 \frac{D^c(\zl)}{c} \left( \frac{\sigma}{c} \right)^4 \left[ 1 - \frac{D^c(\zl)}{D^c(\zs)} \right] \Phi_i  \nonumber\\
&= 1.35 \times 10^{6} \, {\rm s} \, \left( \frac{D^c(\zl)}{1\, {\rm Gpc}} \right) \left( \frac{\sigma}{161 \, {\rm km \, s^{-1}}} \right)^4 \left[ 1 - \frac{D^c(\zl)}{D^c(\zs)} \right] \Phi_i, \label{eq:deltat}
\end{align}
where 
\begin{align}
&\Phi_i  = \frac{1}{2} \left(\textbf{x}_i  - \textbf{y} \right)^2  - \sqrt{ \frac{q (x_{1,i}^2 + x_{2,i}^2)}{1 - q^2} } \,\, \times \nonumber\\
& \left[ \sin{\phi_i} \arcsin{(\sqrt{1- q^2} \sin{\phi_i})} \, + \cos{\phi_i} \, {\rm arcsinh}\left(\frac{\sqrt{1- q^2}}{q} \cos{\phi_i} \right) \right],
\end{align}
where $\sigma$ is the velocity dispersion drawn in Step 4. 
\end{enumerate}	

Figure \ref{fig:mu_deltat_dist} shows the  distributions of $\mu_i$ and $\delta t_i$ corresponding to two prominent images for simulated events before and after applying the detection threshold (SNR=8) in  LIGO-Virgo network.

\subsection{Simulating GW observations}
\label{sec:det_sim}

Appendix~\ref{sec:sampling} describes how we draw random samples of the binary's parameters. Strongly lensed events produced multiple values of the magnification $\{\mu_i\}$ and time delay $\{\delta t_i\}$. Multiply imaged GW signals can be generated by multiplying the original signal with the magnification factor and by applying the lensing time delay  
\begin{equation}
h_{+,\times,\, i}^\mathrm{lens} (f; \bm \lambda) = \sqrt{\mu_i} \ \exp{(\mathrm{i} \, 2 \pi f \delta t_i)} \ h_{+,\times}(f; \bm \lambda), 
\end{equation}
where $h_{+,\times}(f; \bm \lambda)$ are the two polarizations of the original GW signal in Fourier domain corresponding to a set of parameters $\bm \lambda$, $f$ is the Fourier frequency and $\mathrm{i} := \sqrt{-1}$. In practice, we compute different gravitational waveforms by rescaling the luminosity distance $d_L$ by $1/\sqrt{\mu_i}$, at different times $t_0 + \delta t_i$, where $t_0$ is a fiducial reference time. We then project these polarizations on to the Advanced LIGO-Virgo network and compute the optimal signal-to-noise ratio
\begin{equation}
\rho_{i}^{\mathrm{lens}} = 2 \left( \sum_D \int_{f_\mathrm{low}}^{\infty} \, \frac{h_{D, \, i}^{\mathrm{lens}}(f)^2}{S_D(f)} df \right)^{1/2}. 
\end{equation}
Above, the summation is over different detectors, $h_{D, i}^{\mathrm{lens}}(f) := F_{+,\,D}  (\alpha, \delta, \psi) ~ h_{+,\, i}^\mathrm{lens} (f) +  F_{\times,\,D}  (\alpha, \delta, \psi) ~ h_{\times,\, i}^\mathrm{lens} (f)$ denote the observed signal in detector $D$ whose noise has a one-sided power spectral density $S_D(f)$. The antenna patterns of the detector $D$ is denoted as $F_{+,\,D}$ and $F_{\times,\,D}$, which are functions of the source position $\alpha, \delta$ and polarization angle $\psi$. The low-frequency cutoff is chosen to be $f_\mathrm{low} = 20$ Hz. If at least two images have the network SNR $\rho_{i}$ greater than a threshold of 8, we consider them as strong-lensing detections. In our simulation, the fraction of events with more than two detectable images is negligible. We compute the Bayes factors described in Sec.~\ref{bayes_facotr_formulation} using pairs of lensed events as described in Sec.~\ref{sec:astro_sim}. Figure~\ref{fig:injectedredshift_dist} shows the mass and red shift distributions of detectable events.

%
%
%
%
%
%
\bibliography{Lensing_draft}
\end{document}